\RequirePackage[2020-02-02]{latexrelease}
\documentclass[conference]{IEEEtran}

\usepackage{algorithm, algpseudocode}
\usepackage{url}
\usepackage{graphicx,epstopdf}
\usepackage{tabularx}
\usepackage{subfigure}
\usepackage{graphicx}
\usepackage{enumitem}	
\usepackage{listings}
\usepackage{changepage}
\usepackage{pifont}
\usepackage{multirow}
\usepackage{wrapfig}
\usepackage{color,soul}
\usepackage{tikz}
\usepackage{booktabs}
\usepackage[normalem]{ulem}
\usepackage{pifont}
\usepackage{soul}
\usepackage{xspace}
\usepackage{graphicx}
\usepackage{multirow}
\usepackage{multicol}
\usepackage[printwatermark]{xwatermark}
\usepackage{pifont}
\usepackage{epsfig}
\usepackage{tikz}
\usepackage[normalem]{ulem}
\usepackage{multirow}
\usepackage{multicol}
\usepackage{balance}
\usepackage{caption}
\usepackage[bottom]{footmisc}
\usepackage{amsthm}
\usepackage{amsmath,amssymb,amsfonts}
\usepackage{ifthen}

\theoremstyle{definition}

\newcommand{\Paragraph}[1]{\vskip 3pt\noindent\textbf{#1 }}	
\newtheorem{theorem}{Theorem}

\newenvironment{myitemize}%
  {\begin{itemize}
	[leftmargin=0cm,
		itemindent=.3cm,
		labelwidth=\itemindent,
		labelsep=0pt,
		parsep=3pt,
		topsep=2pt,
		itemsep=1pt,
		align=left]
  }%
  {\end{itemize}}

  {\begin{enumerate}
	[leftmargin=0cm,itemindent=.5cm,labelwidth=\itemindent,
		labelsep=0pt,
		parsep=1pt,
		topsep=1pt,
		itemsep=3pt,
		align=left]
  }%
  {\end{enumerate}}

\newcommand{\revmarker}[1]{\sethlcolor{green} \hl{-----Please review until here. Working on content below----}}

\newcommand\numberthis{\addtocounter{equation}{1}\tag{\theequation}}

\newcommand{\ours}{Jarvis}
\newcommand{\run}{Jarvis runtime}

\newcommand{\Proxy}{Control proxy}
\newcommand{\proxy}{control proxy}

\newcommand{\proxies}{control proxies}
\newcommand{\client}{data source}
\newcommand{\Client}{Data source}
\newcommand{\server}{stream processor}

\newcommand{\datapart}{data-level partitioning}
\newcommand{\DataPart}{Data-level Partitioning}

\newcommand{\p}{load factor}

\newcommand{\algo}{StepWise-Adapt}

\newcommand{\allspconfig}{All-SP}
\newcommand{\allsrcconfig}{All-Src}
\newcommand{\filtersrcconfig}{Filter-Src}
\newcommand{\bestopconfig}{Best-OP}
\newcommand{\coarsedp}{LB-DP}
\newcommand{\modelbased}{LP only}
\newcommand{\modelagnostic}{w/o LP-init}
\newcommand{\qone}{S2SProbe}
\newcommand{\qtwo}{T2TProbe}

\newcommand{\pingmesh}{Pingmesh}
\newcommand{\dnnlog}{LogAnalytics}

\usepackage{cite}
\usepackage{graphicx}
\usepackage{textcomp}
\usepackage{xcolor}
\usepackage{comment}
\usepackage{fancyhdr}

\makeatletter
\def\footnoterule{\kern-3\p@
  \hrule \@width 3.5in \kern 2.6\p@} 
\makeatother
\newcommand{\copyrightnotice}[1]{{%
  \renewcommand{\thefootnote}{}
  \footnotetext[0]{#1}%
}}

\begin{document}

\title{Jarvis: Large-scale Server Monitoring with\\ Adaptive Near-data Processing}

\author{\IEEEauthorblockN{Atul Sandur\IEEEauthorrefmark{1}, ChanHo Park\IEEEauthorrefmark{2}, Stavros Volos\IEEEauthorrefmark{3}, Gul Agha\IEEEauthorrefmark{1}, Myeongjae Jeon\IEEEauthorrefmark{2}}
\IEEEauthorblockA{\IEEEauthorrefmark{1}University of Illinois at Urbana-Champaign~~ \IEEEauthorrefmark{3}Microsoft Research~~
\IEEEauthorrefmark{2}UNIST}}


\maketitle
\copyrightnotice{\copyright2022 IEEE.  Personal use of this material is permitted.  Permission from IEEE must be obtained for all other uses, in any current or future media, including reprinting/republishing this material for advertising or promotional purposes, creating new collective works, for resale or redistribution to servers or lists, or reuse of any copyrighted component of this work in other works.}
\renewcommand{\headrulewidth}{0pt}
\chead{\textbf{\textit{\Large Final version published in IEEE ICDE 2022 and received \textcolor{red}{Best Paper Award}. It can be found at \url{https://ieeexplore.ieee.org/document/9835523}}}}


\thispagestyle{fancy}
\pagestyle{plain}

\renewcommand\abstractname{Abstract}
\begin{abstract}

Rapid detection and mitigation of issues that impact performance and
reliability is paramount for large-scale online services.  For real-time
detection of such issues, datacenter operators use a \server{} and analyze
streams of monitoring data collected from servers (referred to as \client{} nodes) and their hosted
services.  The timely
processing of incoming streams requires the network to transfer massive
amounts of data, and significant compute resources to process it.
These factors often create bottlenecks for stream analytics.

To help overcome these bottlenecks, current monitoring systems employ near-data processing by either computing an optimal query partition based on a cost model or using model-agnostic heuristics. Optimal partitioning is
computationally expensive, while model-agnostic heuristics are iterative
and search over a large solution space. We combine these approaches by
using model-agnostic heuristics to improve the partitioning solution from a
model-based heuristic. Moreover, current systems use operator-level
partitioning: if a \client{} does not have sufficient resources to execute
an operator on all records, the operator is executed only on the \server{}.
Instead, we perform data-level partitioning---i.e., we allow an operator to
be executed both on a \server{} and \client{}s.

We implement our algorithm in a system called \ours{}, which enables quick
adaptation to dynamic resource conditions. Our
evaluation on a diverse set of monitoring workloads suggests that \ours{}
converges to a stable query partition within seconds of a change in node
resource conditions.  Compared to current partitioning strategies, \ours{}
handles up to 75\% more data sources while improving throughput in
resource-constrained scenarios by 1.2-4.4$\times$.

\end{abstract}

\begin{IEEEkeywords}
analytics, stream processing, server monitoring, near-data, edge analytics, query partitioning, query refinement
\end{IEEEkeywords}

\section{Introduction}
\label{sec:Introduction}


Today's datacenters use thousands of servers to host large-scale services,
such as web search, database systems, and machine learning (ML) pipelines,
for millions of users. Operating these services with high availability
requires that in order to restore normal service operation, datacenter operators quickly detect performance and reliability
issues\cite{sonata,helios,google-log-analysis} such as high network latency, disk failures, and service outages from
software bugs~\cite{fingerprint,
pingmesh,netbouncer,deepview,slack-observability,autoscaling, m3}. 

\begin{figure}
   \centering
     \includegraphics[width=0.48\textwidth]{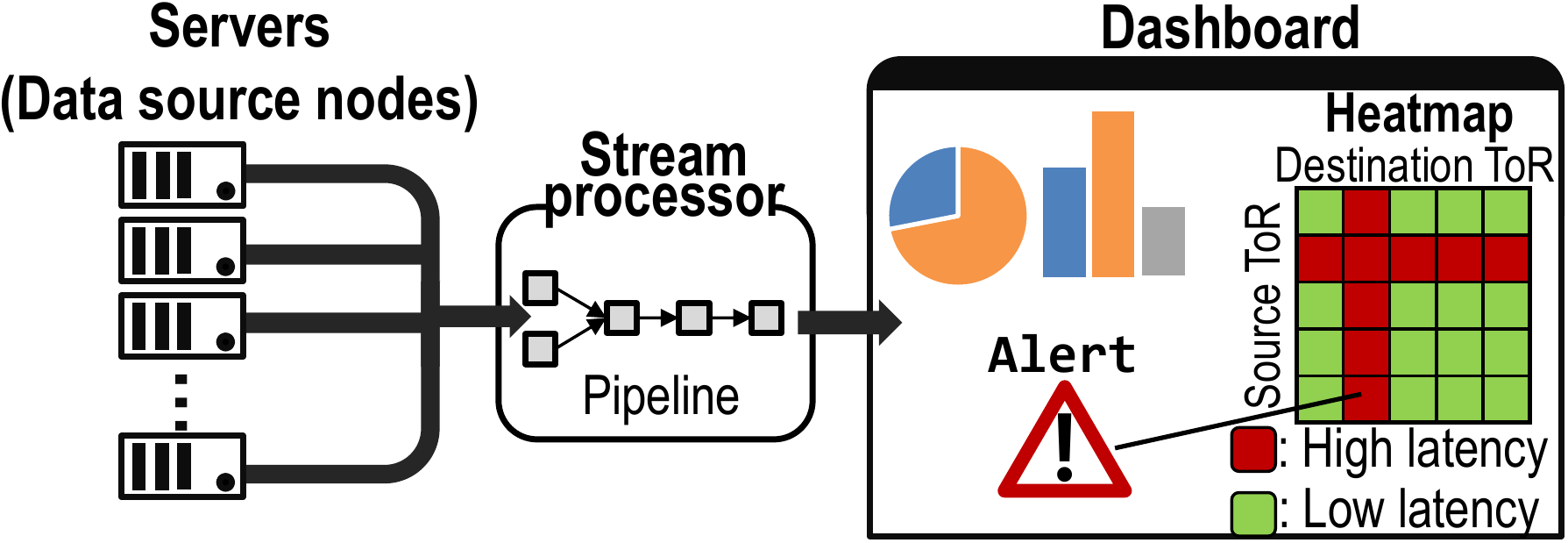}
     \caption{Overview of a datacenter monitoring system.}
     \label{fig:arch}
    \vspace{-0.5cm}
\end{figure}

Datacenter operators deploy a dedicated monitoring system that 
analyzes real-time events as they are streamed from
datacenter servers to a \server{} (Figure~\ref{fig:arch}). As
the collected data is analyzed, the \server{} visualizes the behavior of the
monitored system via dashboards. This allows the operators to generate alerts when issues impacting service availability are observed. Note that the data streamed includes both service-level application logs
and host-level metrics representing the health of
various system resources. Large-scale monitoring pipelines can generate
up to 10s of PBs per day from hundreds of thousands of servers~\cite{chi,
google-log-analysis}, making the network transfer cost to the remote \server{} a
significant bottleneck. 
Furthermore, processing the data in a timely manner requires large amounts of compute resources, which is
increasingly becoming a burden to the \server{}s.

\Paragraph{Challenges.} Various monitoring pipelines leverage available compute 
resources on \client{}s (i.e., server nodes) to process data locally,
effectively reducing the amount of data delivered to the \server{}.
As available resources on the \client{} typically result from over-provisioning
compute resources to handle peak resource demands~\cite{helios, uber-neris-motivation, nutanix-motivation,
resource-overprovision1, resource-overprovision2}, a monitoring query 
may be restricted to run a subset of operations
(e.g., filtering) within a given compute budget to minimize interference with
the hosted services. 
However, such compute budget varies widely across \client{}s and
different time slots in practice (Section~\ref{sec:motivation}). Consequently, prior 
partitioning approaches, which \emph{statically} decide which operations to run on the
\client{}~\cite{everflow,gigascope,jetstream} are either too conservative or, 
alternatively, run out of the assigned compute budget, affecting overall query performance.

To overcome the limitations of static partitioning, Sonata~\cite{sonata} proposes a
\emph{dynamic} approach based on a query cost optimization model. Query operators can be distributed across programmable switches and the \server{} for a wide range of network monitoring applications. 
However, the proposed query partitioning occurs at the operator 
level~\cite{everflow,gigascope,jetstream}, and is modified at runtime by a
central query planner running on the \server{}. Thus, the query planner
deploys to the programmable switch an operator only if its available compute resources 
are sufficient to process \textit{all} of the operator's ingress data.
Unfortunately, such \textit{coarse-grained operator-level} query partitioning is not
effective in scenarios where available compute resources are highly
constrained, as is the case with \client{}s in monitoring systems. Furthermore, 
solving an expensive optimization using an accurate query cost profile and a centralized planner is unsuitable 
for making frequent decisions needed when \client{}s exhibit fast changing 
resource conditions.

\Paragraph{Our proposal.} We propose \ours{}, a new monitoring engine that
targets large-scale systems generating \textit{high-volume data streams}.
For the query workload on each \client{}, \ours{} identifies a fine-grained \emph{\datapart{}} strategy by controlling the amount of data
processed by each query operator, namely \emph{load factor}. This partitioning strategy leverages a model-based technique using query operator costs to estimate the initial load factors and then iteratively adapts the estimates by monitoring the query's execution. \ours{} can significantly reduce network data transfers while utilizing the limited and dynamic compute
resources over \client{}s. As \ours{} is implemented in a fully decentralized
manner, it can scale to a large number of \client{}s.

\ours{} introduces novel extensions to the conventional query execution pipeline: \emph{Control proxy} and \emph{Jarvis
runtime}. The control proxy is a light-weight routing logic---associated with
a query operator---that decides ``how many''
incoming records should be forwarded to the associated query operator. At each \client{}, the local
Jarvis runtime interacts with all control proxies within a query in order to identify their state (i.e., idle, congested, or stable). After observing state changes for the
control proxies of the query, the Jarvis runtime refines its plan for data-level
partitioning in order to keep the query execution in each \client{} stable.

\Paragraph{Results.} We have implemented \ours{} 
and evaluated it with monitoring queries on host-level network latency metrics and application logs. The results show that \ours{} enables a \server{}
node to handle up to 75\% more \client{}s while improving query throughput by
up to 4.4$\times$ over the state-of-the-art partitioning strategies.
Moreover, \ours{} converges to a stable query configuration within seven
seconds of a resource change occurring on \client{}.


\section{Background \& Motivation}



\subsection{Monitoring Systems Overview}
\label{sec:background}

Our work has been motivated by existing large-scale server
monitoring systems~\cite{pingmesh,helios,deepview,fingerprint}. 
We describe two specific scenarios that guide the design of \ours{}:

\begin{myitemize}
\item \textit{Scenario 1:} Network engineers
    deploy Pingmesh~\cite{pingmesh} agents on
    datacenter nodes to collect network latency between node pairs. A web search
    team uses Pingmesh to monitor network health of their latency-sensitive
    service and generate an alert if more than a predefined proportion of
    hosting nodes (e.g., 1\%) have probe latencies exceeding a threshold such as
    5~ms~\cite{pingmesh}.

\item \textit{Scenario 2:} A log processing system, Helios~\cite{helios},
    enables live debugging of storage analytics services such as
    Cosmos~\cite{cosmos}. A bug in a cluster resource manager may lead to service
    resources being under-provisioned. To temporarily mitigate performance
    degradation, several TBs of log streams~\cite{helios} are processed
    quickly to identify impacted tenants. Their latency and CPU/memory
    utilization data are queried to predict resource needs and make scaling
    decisions.
\end{myitemize}

These scenarios rely on a diverse set of computations. Scenario 1 requires
processing of periodically generated metrics composed of structured numerical
data; queries consist of computations specific to numerical data, such as
filtering and aggregation. On the contrary, Scenario 2 requires processing aperiodically generated logs composed of unstructured strings; queries
accompany computations for string processing such as parsing, splitting, and
search.


This diverse set of computations rely on various streaming primitives, which are
computationally different and whose costs are highly dynamic depending on the
input data. For example, filter (F) drops uninteresting records by applying
predicates on each record and typically requires little compute resources. 
Grouping (G) organizes records by key fields, requiring key
lookups in a hash table. Join (J) joins an input stream with a static table, 
requiring key lookups on the table while joining two inputs into an
output stream~\cite{tersecades}.
Unlike filter, both grouping and join operators are expensive due to the
irregular access patterns of hash table lookups. Their cost depends on
the hash table size, which corresponds to the group count and the static
table size, respectively. 
Map (M) performs user-defined transformations on
the input (e.g., parsing and splitting text logs) and its cost depends on the
transformation logic. We present resource usage characteristics of these
primitives in detail in~\cite{jarvis-extended}.

\begin{figure}[t]
  \centering
    \includegraphics[width=3.3in]{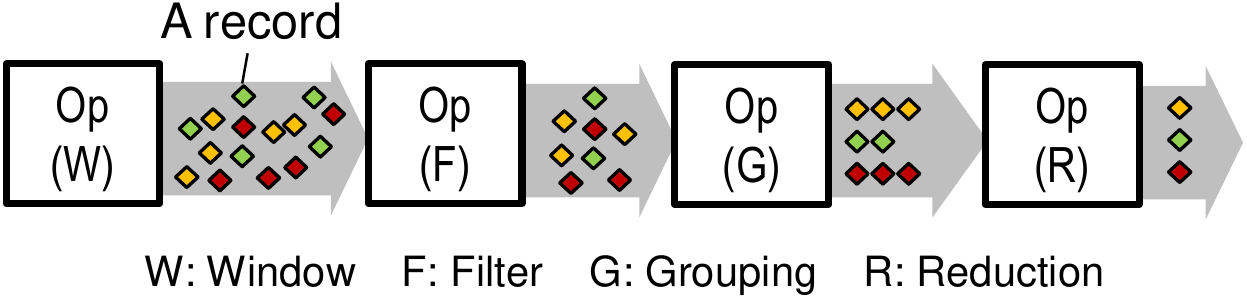}
    \caption{DAG of stream operations for the query in Listing~\ref{lst:motivating-query}.}
    \label{fig:pipeline_tor}
    \vspace{-0.5cm}
\end{figure}

\lstset{
	basicstyle=\fontsize{8}{8}\selectfont\ttfamily,
	xleftmargin=1.0ex,
	framexleftmargin=1.0ex,
	frame=tb, 
	breaklines=true,
	captionpos=b,
	label=list:correct-spi,
	}
\begin{lstlisting}[caption={A temporal query for server-to-server latency probing in every 10-second fixed-size window.}, label={lst:motivating-query}]
/* 1. create a pipeline of operators */
query = Stream
.Window(10_SECS)
.Filter(e => e.errCode == 0)
.GroupApply((e.srcIp, e.dstIp))
.Aggregate(c => c.avg(rtt), c.max(rtt), c.min(rtt))
/* 2. Execute the pipeline */
Runner r( /* config info */ );
r.run(query);
\end{lstlisting}

\Paragraph{Programming model.} We adopt a declarative programming
model~\cite{flink, sparkstreaming} to define a monitoring query. In this
programming model, a query can be generally expressed as a directed acyclic
graph (DAG). In a DAG, the vertices denote stream operations which transform
their input records and emit output records to downstream operations, and the
edges denote dataflow dependencies---i.e., how data flows between the operations. Listing~\ref{lst:motivating-query} shows a real query on the \pingmesh{} data
described declaratively. Figure~\ref{fig:pipeline_tor} illustrates the DAG for 
this \pingmesh{} query. 


\subsection{Adaptive Query Partitioning in Query Engines}
\label{sec:motivation}

\subsubsection{Query Partitioning} Consider the volume of data transferred over the network observed in Microsoft's \pingmesh{} trace. Each record in the trace corresponds to a single probe
message for a server pair. A record is 86B in size, including timestamp (8B),
source IP address (4B), source cluster ID (4B), destination IP address (4B),
destination cluster ID (4B), round trip time in microseconds (us) (4B), and
error code (4B).
Now, assuming that a datacenter consists of 200K servers and each server
probes 20K other servers with a probing interval of 5
seconds~\cite{pingmesh}, the data generation rate is estimated to be
$\sim$512.6~Gbps. Transferring such a high volume of data would strain network
capacity and delay the query execution. Our observations are corroborated by many existing monitoring scenarios which exhibit \textit{high traffic volume}~\cite{chi, gandalf, m3} due to a large number of \client{}s and diverse data streams.

Data synopsis techniques (e.g., sampling, histograms, and sketches)~\cite{synopsis-berkeley} reduce the data sent to the stream processors at the cost of query output accuracy loss~\cite{continuous-sampling-streams, mergeable-summaries}. Despite
their effectiveness, any loss in the query output accuracy may be
undesirable for monitoring tasks (e.g., anomaly detection) wherein
anomalies are hard to detect before their samples are fully
processed~\cite{omnimon}. Data corresponding to anomalies are typically
infrequent, and hence the lossy query output could lead to undetected anomalies that
affect server reliability and additional costs for anomaly
detection~\cite{thread-alert-fatigue}. 


An alternative way to alleviate the network bottleneck is data processing
\emph{near the \client{}}. This approach is motivated by the observation that a monitoring 
query typically consist of a pipeline of operators that incrementally
reduce the data volume. For instance, the query illustrated in Figure~\ref{fig:pipeline_tor}
first filters records and then aggregates them into a few statistical
numbers. For brevity, we refer to the technique of offloading a part of query 
execution near the \client{} as \emph{query partitioning}. Our analysis in 
Section~\ref{sec:sampling-vs-jarvis} demonstrates that query partitioning
can achieve network transfer reduction rates similar to state-of-the-art data
synopsis techniques without impacting the query accuracy. 



\subsubsection{Adaptive Query Partitioning} In order to maximize network transfer reduction rates,
query partitioning schemes should adapt to changes in resource availability and
compute resource demands. 

\Paragraph{Resource availability.} Resource demands of foreground services change dynamically over time. Therefore, unused compute resources on \client{} nodes
available for monitoring queries exhibit temporal variability. This is a
common characteristic of \client{} nodes as they usually host services whose workloads change over time. As a consequence, several large-scale web services (e.g., Alibaba and
Wikipedia) and ML inference services experience bursty request loads in the order of minutes~\cite{alibaba-bursty-workload,
wikipedia-bursty-workload, barista-predictive-analytics,
mark-ml-inference-atc}. 
These services require variable amount of compute resource to meet their SLAs despite changing request loads. Subsequently, query
partitioning decisions need to be made promptly \emph{at runtime} on
\emph{each \client{}}, to be compatible with dynamic resource conditions
available to exploit on each \client{}.


\Paragraph{Resource demands.} 
Resource demands for each \client{} node can vary over time. The root cause is
anomalous behavior, which may cause monitoring data distributions to 
change. Service failures in the datacenter can generate traffic bursts in error log volumes on each \client{}, until the failure is mitigated. Similarly, several real-world log analytics systems report high temporal variability in the record count of log streams, even across one-minute time windows~\cite{chi, data-variable-minutes-dcs, helios}.
In our \pingmesh{} example, network issues can cause spikes 
in server probe latencies, whose duration may range 
between 40 and 60 seconds. 

Changes in the data distribution impacts query 
resource usage characteristics. The output data rate of an operator is a function of the operator's input data distribution and subsequently determines
resource usage of downstream operators. To illustrate, in 
Figure~\ref{fig:pipeline_tor}, the F operator performs filtering and leads to the first data 
volume reduction from the input stream. Since erroneous (or high-latency) records are 
usually not dominant, the operator happens to drop only a small fraction of records. A 
majority of records are then to be processed by G+R operator, increasing compute cost for the query. 
However, if we were to have network issues for inbound or outbound traffic for some servers, 
there would be a high degree of data reduction in F operator, thus lowering resource usage of 
the G+R operator.

Resource demands are also diverse across \client{} nodes due to variable data generation. 
For instance, in our \pingmesh{} example, a subset of servers are configured to probe a larger set of 
peers to cover a larger network range on behalf of other servers in the same ToR switch. This phenomenon
causes highly variable data rates across \client{s}. Our analysis shows that 58\% of the \client{} nodes generate 50\% or lower of the highest rate---details can be found in~\cite{jarvis-extended}.

\Paragraph{Summary.} The ability to quickly adjust query partitioning plans
(i.e., \emph{query refinement}) would reduce the time duration for which queries
either over-subscribe or under-subscribe available compute resources. 
Over-subscription can lead to interference with foreground services on
the \client{}, while under-subscription loses an opportunity to
further reduce outbound network traffic. We seek to develop an approach which not only  reduces the network traffic effectively, given compute budget on each \client{}, but also performs fast query refinement (in the order of seconds) upon a change in the resource availability or resource demands.


\section{Query Partitioning: Definition and Insights}
\label{sec:drawbacks}
In this section, we define the query partitioning problem and then shed light on efficient query partitioning strategies.

\subsection{Definition and Complexity}
Among prior approaches on query partitioning, operator-level partitioning~\cite{everflow,gigascope,jetstream} is one of the most widely used
approaches. Given a query DAG, an operator-level partitioning plan splits operators into those that can be 
executed on the \client{} and those
that require remote execution on the \server{}. The DAG is modified to capture the split using a \emph{boundary 
operator} such that the \client{} executes only the operators
prior to the boundary operator (including itself) in the topological order of the DAG. 

\begin{table}
\begin{tabular}{ |p{0.8cm}|p{7cm}|  }
\hline
 \multicolumn{2}{|c|}{\textbf{Query Partitioning Problem}} \\
 \hline
 $M$ & Number of operators in the query.\\
 $N_d$ & Number of \client{} nodes.\\
 $\vec{b}$ & Vector with an entry $b_i$ indicating the boundary operator of $i^{th}$ \client{}.\\
 $x_{ij}$ & Indicates if the boundary operator of $i^{th}$ \client{} is $j$.\\
 $rc_j$ & Processing cost on \server{} for boundary operator $j$.\\
 $T_{l}(i,b_{i})$ & Local compute cost for operators 1 to $b_{i}$ on $i^{th}$ \client{}.\\ 
 $T_{r}(i,b_{i})$ & Network transfer cost for output of boundary operator $b_{i}$ along with compute cost of executing operators $b_{i}+1$ till $M$ on \server{}.\\
 \hline
\end{tabular}
\caption{\label{tab:notation-query-partition} Variables in the query partitioning problem.}
\vspace{-0.5cm}
\end{table}

\Paragraph{Problem.} Table~\ref{tab:notation-query-partition} summarizes the variables used to define our problem. Let us consider the vector of boundary operators $\vec{b}$ for a given query. Our goal is to find $\vec{b}$ which minimizes the number of operators sent to \server{} for remote execution, without sacrificing query processing time as a result of the operators executing locally on \client{}, as follows: \vspace{-0.2cm}

\begin{align*}
    & \displaystyle \min_{\vec{b}} \sum_{i=1}^{N_{d}}\sum_{j=1}^{M} rc_j x_{ij} \numberthis
    \label{eq:optimization-NP}
\end{align*}
subject to $T_{l}(i,b_{i}) <= T_{r}(i,b_{i})\ \forall b_{i}>0, i\epsilon[1,N_{d}]$. 

We incentivize executing operators on \client{}s, so the partitioning costs are ordered as $rc_1>rc_2>,...>rc_M$.
Unfortunately, solving this partitioning problem is challenging.

\begin{theorem}
Query partitioning problem in Eq.~\ref{eq:optimization-NP} is NP-hard.
\end{theorem}

\begin{proof}
The key idea of the proof is that the generalized assignment problem
(GAP)~\cite{gap} can be reduced to a special case of our query partitioning
problem in polynomial time. Therefore, if we have an algorithm that can
minimize the number of operators sent to the \server{} without sacrificing
query processing time, then we can obtain an optimal solution to GAP. Since
GAP is NP-hard, our problem is also NP-hard. Refer to~\cite{jarvis-extended}
for the details of the proof.
\end{proof}

Query partitioning involves determining the
boundary operator for the query instance running on each \client{}. Due to \server{} resources being shared across \client{}s, we need to jointly identify boundary operators across \client{} nodes which is exponential in the number of nodes. Given that
resource conditions can change in the order of minutes
(Section~\ref{sec:motivation}), quickly computing a new partitioning
plan is critical. Our setup consists of a \server{} monitoring up to 250 \client{}s, making it prohibitively expensive to compute a new plan for all the \client{}s.

\subsection{Insights}
\label{sec:insights}
\Paragraph{Combining model-based and model-agnostic techniques for query
refinement.} To refine query plans with high efficiency, we exploit a greedy
and embarrassingly parallel heuristic, which reduces the outbound network traffic. 
We utilize a combination of model-based and model-agnostic techniques 
to allow each \client{} node to make fast and effective query refinement decisions \emph{independently}.

This model-based technique quickly finds a new partitioning plan based on
online and fine-grained profiling of each operator's compute cost. However,
such profiling might be inaccurate if the compute resources available for
executing operators are insufficient for accurate profiling. Thus, we apply a
model-agnostic process to iteratively fine-tune the output plan produced by
the model-based approach. The proposed heuristic is simple,
computationally tractable, and suggests a reasonable first attempt.

\begin{figure}[t]
\centering
\includegraphics[width=\linewidth]{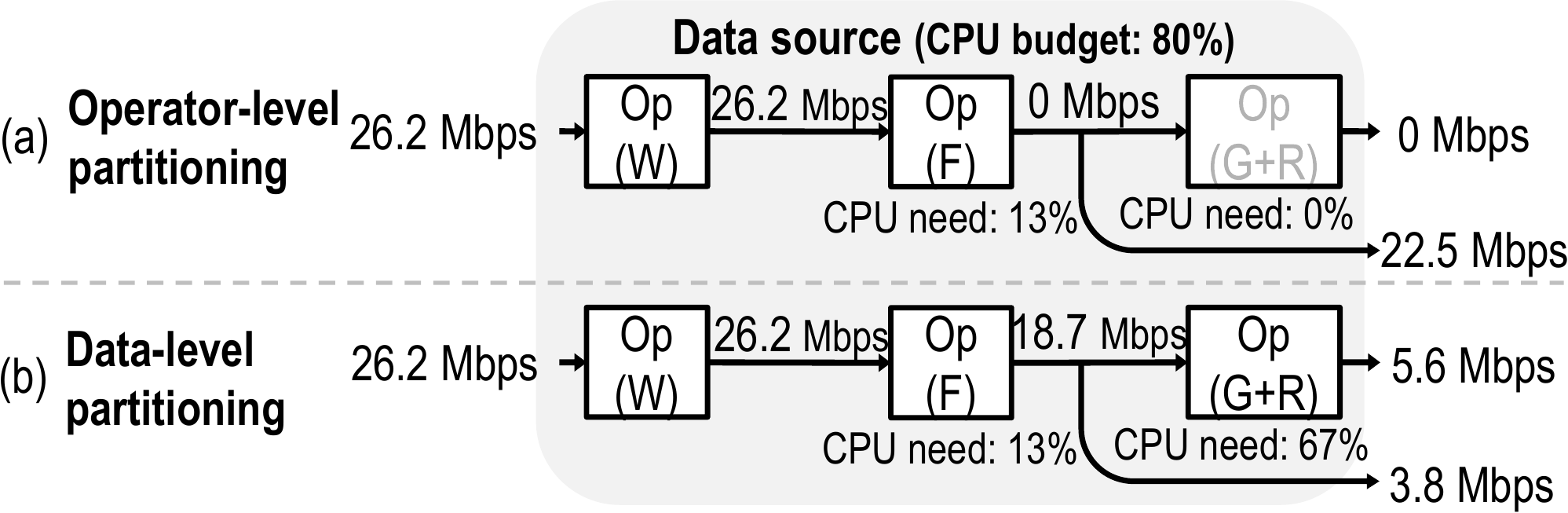}
\caption{\label{fig:data-partition-explain} Coarse-grained operator-level vs. fine-grained data-level partitioning on \client{} with 80\% CPU budget. G+R requires 80\% CPU to process all output data from F operator.}
\vspace{-0.3cm}
\end{figure}

\Paragraph{Data-level query partitioning.} The operator-level query
partitioning could be undesirable when attempting to run a computationally
expensive operator like grouping within limited compute resources provided by
\client{}. Instead, \ours{} adopts \datapart{} to allow an operator to
process a fraction of its input records on \client{} and drain the rest for
remote processing. This \datapart{} is fine-grained and can improve the
overall utilization of limited compute resources on each \client{} node.

We highlight the effectiveness of \datapart{} by performing an empirical study using the query in 
Figure~\ref{fig:pipeline_tor} on a real-world \pingmesh{} trace. Our experiment runs the query on a 
\client{} node with compute budget set to 80\% of a single 2.4~GHz CPU.
Figure~\ref{fig:data-partition-explain}(a) shows that operator-level 
partitioning cannot execute the costly G+R operator entirely within compute budget, because the F 
operator drops only a small portion of input records. This leads to network traffic as high as 22.5 Mbps, 
which is close to the input rate, while not fully utilizing the compute budget. On the contrary,
Figure~\ref{fig:data-partition-explain}(b) shows that the operator G+R can utilize the compute budget 
fully and process 83\% of its input under the \datapart{}, resulting in total network traffic of 9.4 
Mbps---i.e., 2.4$\times$ lower over the operator-level partitioning.

\section{Jarvis}
\label{sec:jarvis}
We discuss the design of \ours{} and our proposed \datapart{} heuristic which works in a decentralized manner.


\subsection{Design Overview}
\label{sec:architecture}

\begin{figure}
\centering     
\subfigure[]{\label{fig-design:b}\includegraphics[width=42mm]{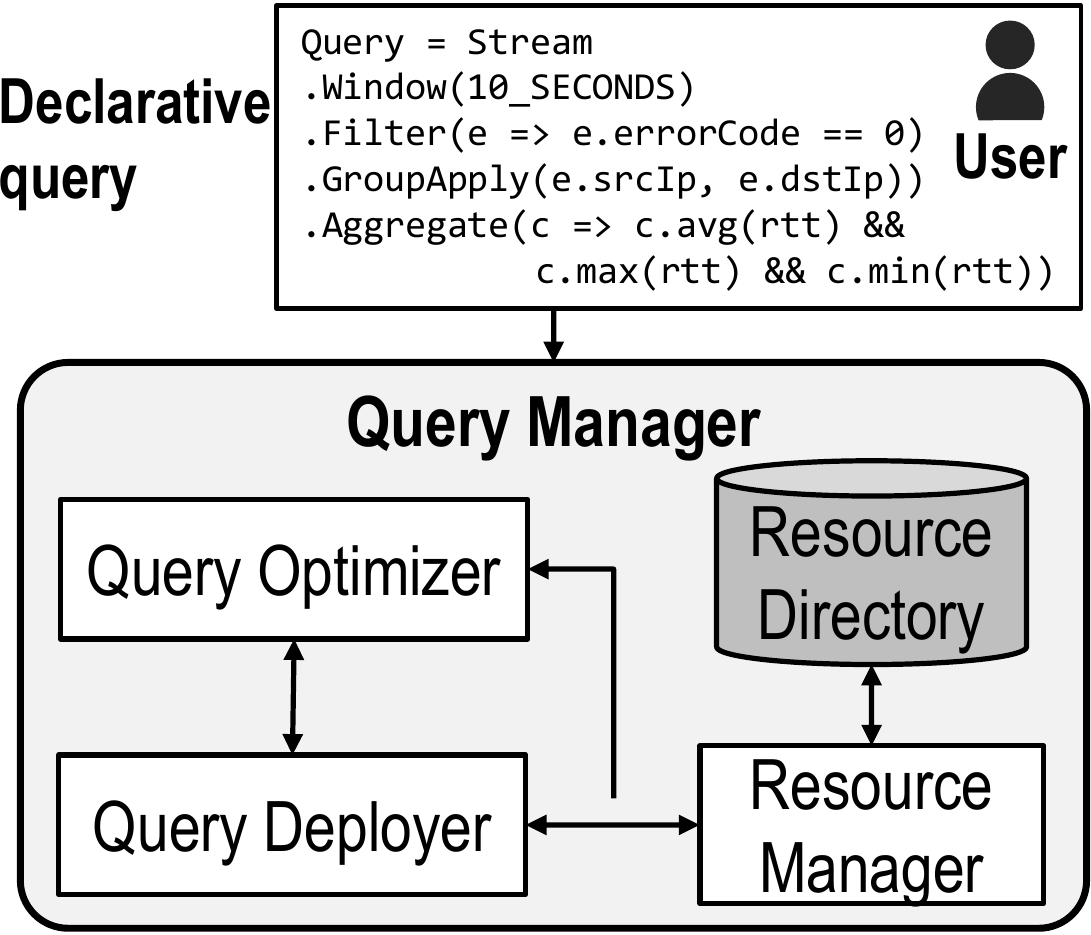}}
\subfigure[]{\label{fig-design:a}\includegraphics[width=42mm]{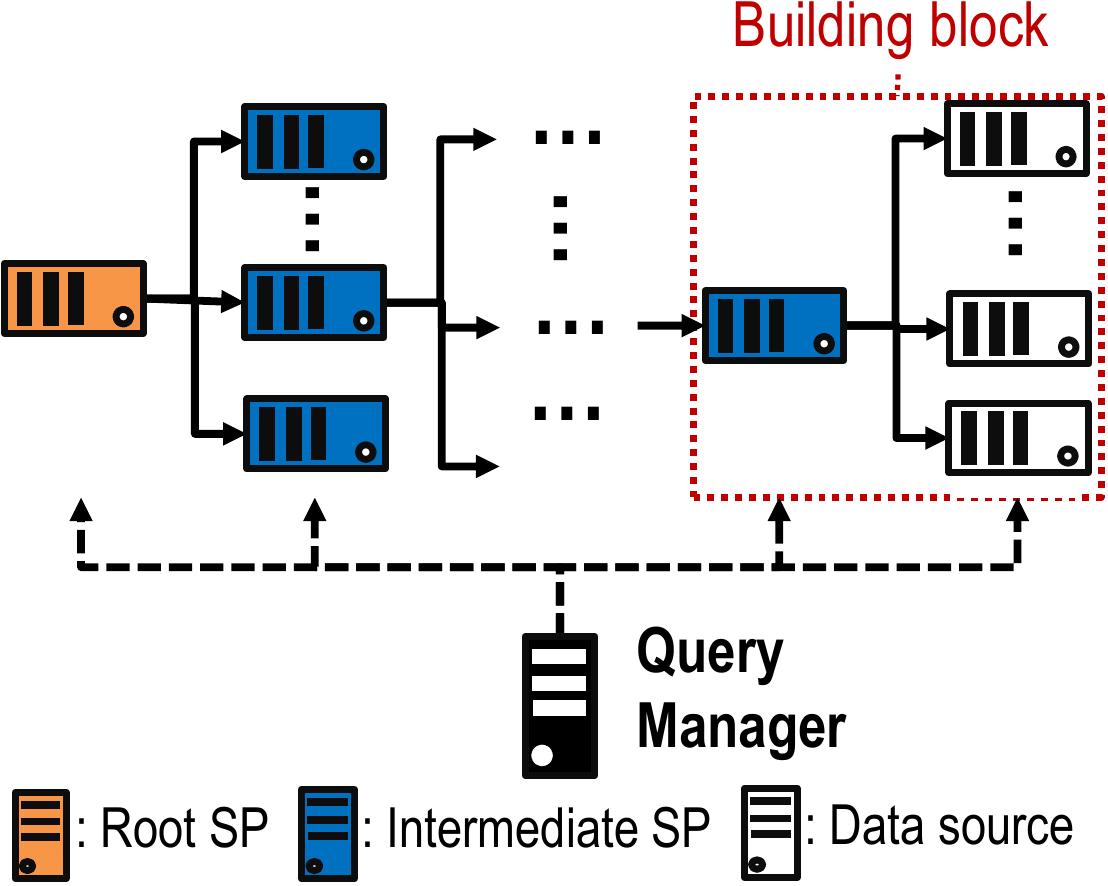}}
\caption{\label{fig:arch-monitoring-scale} (a) \ours{} query manager, and (b) Monitoring pipeline architecture. SP refers to stream processor. $L_{0}$ to $L_{H}$ refer to the hierarchy levels in a tree topology of height $H$.}
\vspace{-0.35cm}
\end{figure}

Figure~\ref{fig:arch-monitoring-scale}(a) shows the key system components
involved in the \emph{query manager} when a user submits a query to \ours{}. The
query manager includes the \emph{resource manager}, which maintains in the \emph{resource directory} the list of \client{} and \server{} nodes along with their network topology. The \emph{query optimizer} uses the query topology
information to generate an optimized physical plan from the logical plan for
the query, as is done in most streaming engines~\cite{flink}. The \emph{query
deployer} compiles the query plan into an executable code and deploys it on
\ours{} running on each node. Figure~\ref{fig:arch-monitoring-scale}(b) shows the query manager maintaining the resource node topology and deploying the query. Our work focuses on changing the query optimizer to generate a \ours{}-friendly query plan and runtime optimizations after the query is deployed on the nodes.

As shown in Figure~\ref{fig:arch-monitoring-scale}(b), physical resources involved in 
query execution are viewed as a \emph{tree structure}, where leaf 
nodes represent the \client{} and the rest of the nodes represent
the \server{}. Each \client{} node executes the query within its available compute budget and sends the results, along with any pending data that needs to be processed, to its parent \server{} node. The parent node leverages its compute resources to further aggregate the query results for its \client{} nodes. The combination of \client{} nodes and the common parent node constitutes a core building block. 
As multiple core building blocks are present 
in a large-scale monitoring scenario,
we should maintain a collection of intermediate \server{} nodes. The root node aggregates the results from these nodes to compute the final query output. As there is no communication between building blocks, the system can scale better to handle more \client{}s if each core building block is more scalable. 
Hence, for the remainder of this work, we focus on optimizing query execution on a single core building block.



\ours{} ``replicates'' query 
operators across \client{} and \server{} nodes. It realizes data-level partitioning and runtime query refinement, by adding  
two novel 
primitives to the query execution pipeline: (1) \emph{\Proxy{}}, a 
unified abstraction for stream operator and (2) \emph{\run{}}, a 
system runtime that coordinates executions of all \proxies{}. 
\Proxy{} is a light-weight operator bridging two 
adjacent stream operators, which decides ``how many'' records shall
be forwarded to the downstream operator vs. to the ``replicated'' operator on the remote \server{} node. 
The query optimizer in \ours{} adds \proxies{} to the query plan at compile time. 


\begin{figure}
  \centering
    \includegraphics[width=0.48\textwidth]{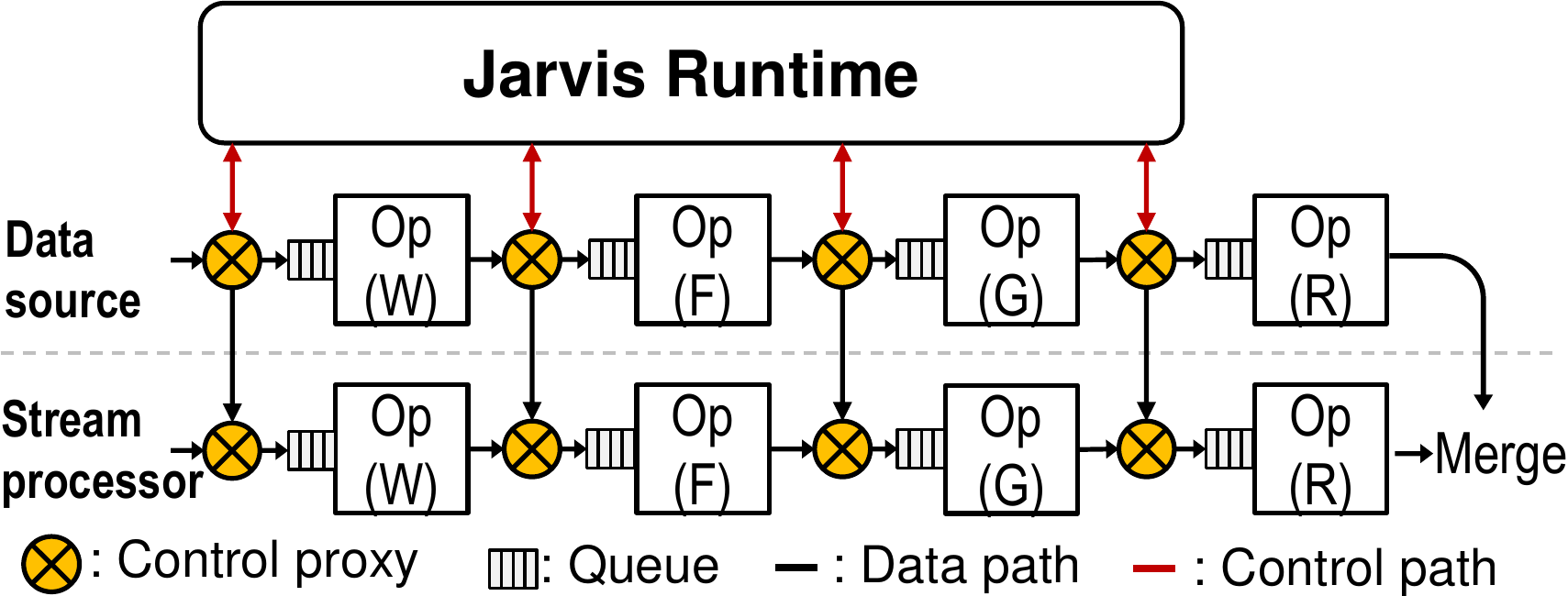}
    \caption{An overview of \proxies{} and \run{}.
    Data path: routing path for incoming data. Control path: interaction between \proxies{} and \run{}.
    }
    \label{fig:jarvis_query}
    \vspace{-0.35cm}
\end{figure}

Figure~\ref{fig:jarvis_query} illustrates the query plan
deployed on \client{} and \server{} for the example query from
Figure~\ref{fig:pipeline_tor}. \run{} deployed at each \client{} configures
\proxies{} to execute a \datapart{} plan for the query. \run{} continually
probes the states of \proxies{} to observe the query state. If the query is
in either idle or congestion state due to changes on resource conditions,
\run{} computes a new \datapart{} plan by reconfiguring the \proxies{}.


\run{} is fully decentralized---i.e., each instance of the query on a
\client{} has a dedicated runtime instance which functions independent of
other query instances and \client{}s. Its interactions with \proxies{} are
also local to the node, requiring no coordination with an external planner or
the \server{}. In essence, our design avoids having a central service
performing a joint optimization across queries and \client{}s~\cite{sonata},
which might be computationally expensive with a large number of \client{}s.


The \datapart{} design in \ours{}
is distinct from conventional backpressure mechanisms. Rate throttling and data 
dropping~\cite{backpressure-spark-streaming, rxjava-backpressure} 
mitigate over-subscription of compute resources at \client{}s, but at the cost of losing accuracy in the query output similar to data synopses. 
Lossless backpressure mechanisms, such as resizing of operator queues~\cite{flink-backpressure,rxjava-backpressure} and operator scaling~\cite{spark-dynamic-scaling-backpressure} have limited effectiveness due to resource constraints on \client{} node. 

\subsection{Query Plan Generation}
\label{sec:query-plan-generation}


\ours{}' query plan generation is built upon the conventional workflow of existing streaming engines~\cite{calcite}. First, the input query is parsed to confirm
its syntax correctness. Then, a logical plan is constructed along with logical optimizations, 
such as constant folding, predicate pushdown. \ours{} inserts a \proxy{} between each of the adjacent stream operators in the logical plan.
Finally, the optimized logical plan is translated to a physical plan for deployment and execution. 
Note that all the above steps are transparent to users.



\ours{} does \emph{not} currently support all operators on
\client{}. Such operators are identified using the following rules:
(\textbf{R-1}) Aggregation operators that are not incrementally updatable,
such as exact quantiles. However, their approximate versions, such as
approximate quantiles~\cite{aomg,prometheus-estimate-percentile}, can benefit
from \ours{}; (\textbf{R-2}) Downstream operators succeeding stateful
operations that require aggregation across multiple \client{}s; (\textbf{R-3})
Stateful joins across streams. Similar to prior work~\cite{sonata}, we note that such operations are expensive, may not reduce outbound data,  
and require processing streams across \client{}s; and (\textbf{R-4}) Multiple physical 
operators per logical operator, useful for parallelizing operator execution
(e.g.~\cite{ds2}). \Client{}s have constrained compute budget, so the benefits of exploiting intra-level operator parallelism would be limited.


The rules are described in a configuration file and can be extended. All rules except R-4 apply also to intermediate \server{}s. This is because R-1 to R-3 consider operations which cannot be incrementally executed or may not result in data reduction from incremental processing. For R-4, however, intermediate \server{}s are dedicated to run monitoring queries, enabling hardware-level parallelism to be exploited to accelerate operator execution. It may appear that R-1 limits queries that can leverage \ours{}. However, a significant number of real-world queries use operators that support incremental updates. For example, Drizzle~\cite{drizzle} has reported that 95\% of aggregation queries
on a popular cloud-based data analytics platform consist of aggregation operators
supporting incremental updates, such as sum and count.

Once the rules above are applied, queries deployed on \client{}s typically consist
of a chain of operators. Hereafter, 
our scope is restricted to such operator pipelines.
Nevertheless, our approach can be extended to handle graphs with split patterns
that may execute on \client{}s---i.e., the output of an operator is an input to
multiple downstream operators.

\subsection{Dynamic Query Refinement}
\label{sec:dynamic-refinement}

\begin{figure}
  \centering
    \includegraphics[width=0.48\textwidth]{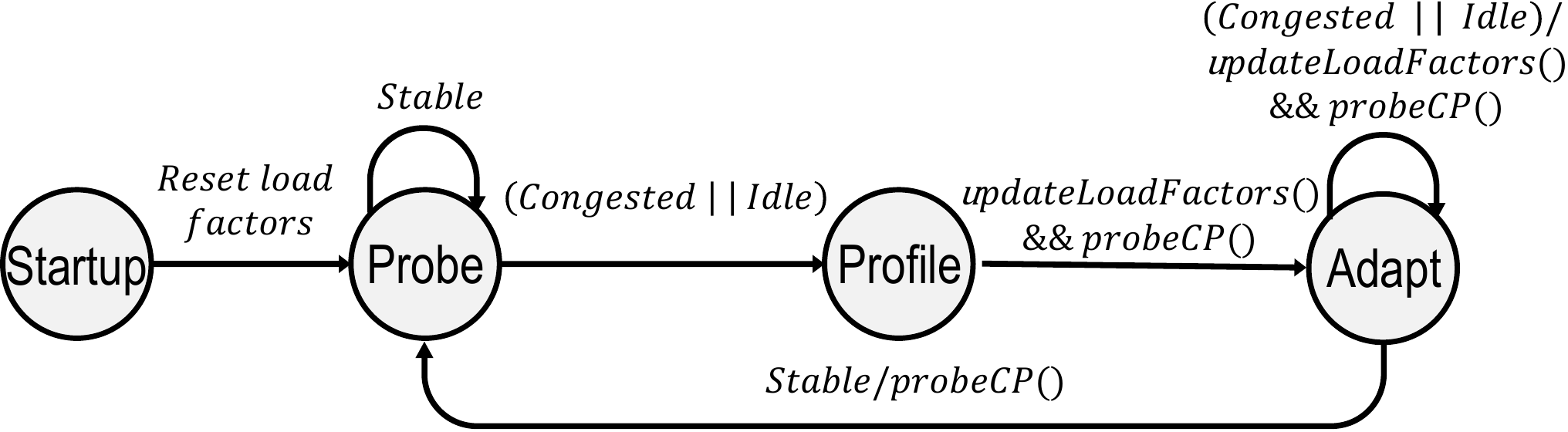}
    \caption{\run\ state machine.}
    \label{fig:jarvis_state_machine}
    \vspace{-0.4cm}
\end{figure}

Once a query is deployed, \run{} steers each
\proxy{} to select a portion $p$ (i.e., 0 $\leq p \leq$ 1) of incoming data
to be forwarded to the downstream operator via the downstream queue. 
Remaining data is drained over the network to be processed by the
\proxy{} associated with the same downstream operator in the \server{}.
Hereafter, we refer to the $p$ value of a \proxy{} as its \emph{load factor}.


\ours{} can continuously refine \p{}s of \proxies{} in the \client{} node to
adjust a query partitioning plan. This query refinement occurs at an epoch
boundary defined by a time interval. During epoch processing, each \proxy{}
monitors its downstream operator to identify the state of the operator to be
one of the following: (\textbf{Congested}) Operator contains more than a
predefined number of pending records, experiencing backpressure;
(\textbf{Idle}) Operator stays empty for longer than a predefined time
duration; and (\textbf{Stable}) Operator is neither congested nor idle. \run{}
collects state information from all \proxies{} at the end of an epoch and
classifies the current \datapart{} plan as \emph{non-stable} if all operators
are idle or at least one operator is congested. Upon identified as
non-stable, \run{} triggers adaptation to bring the query back to the stable
state.


Figure~\ref{fig:jarvis_state_machine} illustrates an overall workflow 
of operational phases in \ours{} to keep stable query executions:


\begin{myitemize}

    \item \textbf{Startup: initialization.} All \p{} values are initialized to
    zero, so all records are processed by \server{}.

    \item \textbf{Probe: normal operation.} At the end of every epoch, \run{}
    executes \texttt{ProbeCP()} function to query all \proxies{} and
    determine their congestion states. It continues to do so until it
    identifies the computation pipeline as congested or idle. At this point,
    the runtime enters \texttt{Profile} phase.

    \item \textbf{Profile: query plan diagnosis.} \ours{} obtains new estimates
    for the following during an epoch: (1)
    compute cost of each operator by executing an operator at a time,
    (2) reduction in the size of input stream by executing each operator, and (3) available
    compute budget for the query. These estimates are used to adapt load factors in the next
    \texttt{Adapt} phase.

    \item \textbf{Adapt: load factor adaptation.} \run{} computes a new \datapart{} plan.
    Initial \p{}s are
    first calculated using the profiling estimates and set for each \proxy{}.
    \run{} executes \texttt{ProbeCP()} to probe the query state and perform iterative
    fine-tuning if necessary, until the computation pipeline is back to stable state.
    At this point, it returns to \texttt{Probe} phase.

\end{myitemize}

Small workload variation in congested state or idle state can trigger a
series of \texttt{Profile-Adapt} phases that put the control system in an
oscillating behavior with small implications on the optimal load factors. To
avoid this undesired behavior, each \proxy{} is configured with a threshold
fraction of pending records in an epoch (\texttt{DrainedThres}) that can be
drained by \proxies{} and tolerated by \texttt{ProbeCP()} without signaling
congested state. Similarly, each \proxy{} is configured with a threshold
fraction of epoch duration (\texttt{IdleThres}) that allows \proxies{} to
stay idle and can be tolerated by \texttt{ProbeCP()} without signaling idle
state.



\subsection{Computing \DataPart{} Plan}
\label{sec:data-partitioning-problem}
Computing a new \datapart{} plan, required for
stabilizing the query execution, is achieved  by solving an optimization
problem to compute initial \p{}s followed by an iterative process for  fine-tuning \p{}s.



\begin{table}
\begin{tabular}{ |p{0.3cm}|p{7.5cm}|  }
\hline
 \multicolumn{2}{|c|}{\textbf{Data-level Partitioning Problem}} \\
 \hline
 $N_r$ & Total number of records injected into the query in an epoch.\\
 $Op_{j}$ & $j^{th}$ operator in the query.\\
 $r_{j}$ & Relay ratio of $Op_{j}$, i.e., ratio of its output 
 to input data size.\\
 $c_j$ & Compute cost of $Op_{j}$ for a single record.\\
 $d_{j}$ & Number of records drained in an epoch at the $j^{th}$ \proxy{}.\\
 $C$ & Compute budget available to the query.\\
 $p_{j}$ & $j^{th}$ \proxy{}'s load factor, i.e., fraction of incoming records to be processed by downstream operator.\\
 $e_{j}$ & Effective load factor for $j^{th}$ \proxy{}, i.e., product of load factors of upstream query operators until $Op_{j}$.\\
 \hline
\end{tabular}
\caption{\label{tab:notation-data-part} Variables in the data-level partitioning problem.}
\vspace{-0.5cm}
\end{table}

\Paragraph{Problem definition.} Table~\ref{tab:notation-data-part} summarizes the variables used to define our problem. Let us consider a
computation pipeline which contains $M$ operators $\{Op_i : 1 \le i \le M\}$.
Here, we choose \p{}s that minimize the total number of drained
records (i.e. $\sum_{i=1}^{M}d_i$) from \client{} given the compute budget
$C$ available to the query: 
\begin{align*}
    \min_{p_{1},p_{2},...p_{M}}
    &\sum_{i=1}^{M}[\prod_{j=0}^{i-1}p_{j}r_{j}](1-p_{i}) \numberthis
    \label{eq:optimization-obj}
\end{align*}
subject to $\sum_{i=1}^{M}[\prod_{j=0}^{i-1}p_{j}r_{j}] p_{i}c_{i} \le C/N_r,$ 
\begin{align*}
    &0 \le p_{i} \le 1,\ 0 \le r_{i} \le 1,\
    c_{i} \geq  0\ \forall\ i\
    \epsilon\ [1,M], p_{0}=1, r_{0}=1
\end{align*}
\noindent Note that $N_r$, $M$, and $C$ are fixed for a problem instance.

Unfortunately, solving the optimization problem is challenging.
First, the formulation is non-convex and hence computationally hard (proof can be found in~\cite{jarvis-extended}.) While it is feasible to enumerate all possible combinations of load factor values across operators, doing so is expensive for online optimization. Second, the formulation assumes certain
conditions, which may not always be satisfied in practice. For instance, to
estimate $c_i$ accurately, each operator needs to be evaluated on a
sufficient number of input records.
Finally, relay ratio $r_i$ can vary non-linearly for certain operations, such as
grouping where $r_i$ is affected by input's grouping key distribution.



\Paragraph{\algo{} algorithm.} \emph{\algo{}}, a novel hybrid algorithm, lies in the
heart of \ours{}' data-level partitioning approach. The algorithm
combines two techniques: (1) a model-based technique which searches for
near-optimal \p{}s based on the modeling assumptions of the \datapart{}
problem defined in Equation~\ref{eq:optimization-obj}, and (2) a
model-agnostic technique that monitors query execution using load factors
obtained from step (1) and fine-tunes them if the query behavior deviates
considerably from stable state---i.e., the available resources are
over/under-subscribed by the query. For fast fine-tuning, step (2) uses a
heuristic inspired by the first fit decreasing (FFD) heuristic for bin packing
problem~\cite{first-fit-bin-pack}, to prioritize \p{} updates of operators
that contribute to higher network traffic reduction.

The first step entails solving the optimization in Equation~\ref{eq:optimization-obj} quickly and efficiently. In doing so, we considered whether a transformation (i.e., change in optimization 
variables) exists to make the objective and constraint functions convex. Based on the performance analysis of optimization solvers on problem formulations resulting from
different transformations (details in~\cite{jarvis-extended}), we settled on a transformation, which yields a linear program (LP).
The transformation is done by
introducing a new optimization variable $e_i$ for the $i^{th}$ \proxy{} where
$e_i=\prod_{j=0}^{i}p_j$. Then, the optimization problem in
Equation~\ref{eq:optimization-obj} can be rewritten as:
\begin{align*}
    \min_{e_{1},e_{2},...e_{M}}
    &\sum_{i=1}^{M}[(\prod_{j=0}^{i-1}r_{j}).(e_{i-1}-e_{i})] \numberthis
    \label{eq:optimization-obj-linear}
\end{align*}
subject to $\sum_{i=1}^{M}[(\prod_{j=0}^{i-1}r_{j}).e_i.c_i] \le C/N_r,$
\begin{align*}
    &0 \le e_{i} \le e_{i-1}\ \forall\ i\ \epsilon\ [1,M],\ e_{0} = 1
\end{align*}

\noindent The rest of conditions on $N_r, M, C, c_i, r_i$ remain the same as in
the original formulation. 

A feasible solution provided by LP solver assumes that operator relay
ratios/costs of operators (i.e., $r_i$ and $c_i$) are fixed and independent
of load factors. However, in case these parameters are unsteady, the solver
provides \p{}s which would either over-subscribe or under-subscribe the
compute budget, making the query execution unstable.
\algo{} takes the second step to address the issue. 

In the second step,
\algo{} observes the query state after it executes an epoch with current
\p{}s of \proxies{} and fine-tunes them based on the priorities of their downstream
operators. 
Operators are assigned \emph{higher} priority based on if they exhibit \emph{lower} data relay ratio. If the query is in the idle state, \algo{} then aims to increase the \p{} of the operator with highest priority first (until its $p=1$). On the contrary, if the query is in congested state, \algo{} then aims to decrease the \p{} for the operator with lowest priority first (until its $p=0$). This approach enables the algorithm to give more resources to operators that potentially result in higher data reduction. When fine-tuning a \p{}, the algorithm executes a binary search over discretized \p{} values to further improve
convergence time. Note that it is possible to use other definitions for operator priority (e.g., priority which considers operator compute cost along with relay ratio), and we leave its investigation for future work.

\subsection{Discussion}
\label{sec:limitations}
\Paragraph{Adaptation.} The effectiveness of \ours{}' adaptation is maintained as 
long as, (1) the epoch duration is large enough to avoid invoking query partitioning
decisions too frequently, and (2) query workload characteristics do not change 
dramatically during \texttt{Profile} and \texttt{Adapt} phases. When setting epoch 
duration to one second, \ours{} requires up to seven seconds (for the evaluated 
workloads) to stabilize a query. The convergence time is acceptable 
for scenarios where resource conditions change in the order 
of minutes (Section~\ref{sec:motivation}).

\Paragraph{Multiple queries.} Multiple queries can run on a \client{} node, with each query having a dedicated \run{} instance. We adopt a fair resource
allocation policy~\cite{minmax} to allocate the compute budget among
competing queries. Running multiple queries is evaluated in
Section~\ref{sec:multiple-nodes-eval}. We leave opportunities for better
resource allocation strategy and operator sharing across
queries~\cite{mqo,sbon} for future work. 

\Paragraph{Fault tolerance.} The system may exhibit a \client{} or \server{} node failure. Checkpointing intermediate state (e.g., intermediate aggregation results of a stateful G+R operation) accumulated by the \client{} for the current window (via the drain path from proxies{}) can enable the \server{} to process remaining data for the current window. Checkpointing can impact network traffic; hence depending on query requirements, we could determine when it should occur based on observed events (e.g., anomalous data in the stream) or a configurable frequency parameter. Likewise, checkpointing the intermediate state of \server{} node can enable the \client{} to replay records produced after the last successful checkpoint. More details on our ongoing efforts to handle failures in \ours{} can be found in~\cite{jarvis-extended}.

\section{Implementation}
\label{sec:implementation}
We implemented \ours{} using Apache MiNiFi~\cite{minifi}, a lightweight query execution runtime, on the \client{} side and Apache NiFi~\cite{nifi} on the \server{} side. RxJava is used to implement query computation pipelines within NiFi/MiNiFi custom processors. Kryo serialization framework~\cite{kyro} is used for transferring data over the network between \client{} and \server{} nodes. Below, we highlight major implementation issues that \ours{} addresses, with a comprehensive discussion provided in~\cite{jarvis-extended}.

\Paragraph{Integration with existing streaming engines.} \ours{} can be integrated with existing query engines on \server{} side. On the \client{} with limited compute resources, unlike stream processors which require dedicated server resources (e.g., Flink~\cite{flink}), it is preferable to run a lightweight dataflow runtime such as MiNiFi. MiNiFi allows us to design and deploy custom dataflows, so it is easy to incorporate \run{} and \proxy{} during query compilation (Section~\ref{sec:query-plan-generation}). MiNiFi agents send data to NiFi on \server{} and we leverage NiFi’s integration already available with existing query engines; for example, Flink provides NiFiSource and NiFiSink connectors to exchange data with NiFi~\cite{flink-nifi-connector}. 
Query instances on \server{} only require adding \proxy{} to the dataflow (Figure~\ref{fig:jarvis_query}), which can be implemented using output tags~\cite{flink-side-output} for splitting the incoming stream to multiple downstream operators.

\Paragraph{Accurate query processing.} Streams from multiple \client{}s are processed on the \server{} side. Accurate stream processing introduces two requirements. 

First, multiple streams need to be consumed via correct merging of watermarks from each stream to indicate time progress. We implement this requirement by leveraging the methodology used by Flink~\cite{flink-multiple-streams}. Each operator advances its time based on the minimum of all its incoming input streams’ event times. Since each \proxy{} on the \client{} generates an additional stream for drained records, incoming watermarks need to be replicated by \proxy{} for the drained path to reflect time progress. 

Second, records emitted by \client{}s need to be routed to the right operator on \server{} for further processing. In doing so, \proxy{} attaches an identifier for the operator on \server{} that should receive records for further processing. While stateless operators on the \client{} can relay their output to the downstream query operator on \server{}, stateful operators relay output to the corresponding operator on \server{}, for merging the accumulated state on \client{} with the state on \server{}. 
\section{Evaluation}
\label{sec:evaluation}


\subsection{Methodology}
\label{sec:exp-methodology}


\Paragraph{Testbed setup.} We deploy our \client{} on Amazon EC2 t2.micro
nodes, each with one 2.4~GHz Intel Xeon E5-2676 core and 1~GB RAM. We use a
larger Amazon EC2 t2.medium node with two 2.4~GHz Intel Xeon E5-2686 cores
and 4~GB RAM as \client{} for the experiment of multiple queries. A \server{}
instance is deployed on an Amazon EC2 m5a.16xlarge node with 64 2.5~GHz AMD
EPYC 7000 cores and 256~GB RAM. All nodes run Ubuntu 16.04.


We conduct our experiments on two types of setups: (1) a single \client{}
node connected to a single \server{} node to evaluate the performance of query
partitioning and refinement strategies in \ours{}, and (2) multiple \client{}
nodes (up to 250) connected to a single \server{} node to evaluate \ours{} as
we increase the monitoring scale.


\Paragraph{Performance metrics.} We measure \textit{query processing
throughput} in Mbps (megabit per second) with a latency bound of 5 seconds,
\textit{epoch processing latency} in seconds, and \textit{convergence
duration} in number of epochs after resource conditions change. Performance
results are obtained after running three minutes for the system warm-up.


\lstset{
	basicstyle=\fontsize{8}{8}\selectfont\ttfamily,
	xleftmargin=1.0ex,
	framexleftmargin=1.0ex,
	frame=tb, 
	breaklines=true,
	captionpos=b,
	label=list:correct-spi,
	}
\begin{lstlisting}[caption={A temporal query for ToR-to-ToR latency probing. $m$ is a table to map server IP address to its ToR switch ID.}, label={lst:tor-level-count-probe}]
Stream
.Window(10_SECS).Filter(e => e.errCode == 0)
.Join(m, e => e.srcIp, m => m.ipAddr, 
     (e,m) => (e, srcTor=m.torId))
.Join(m, e => e.dstIp, m => m.ipAddr, 
     (e,m) => (e, dstTor=m.torId))
.GroupApply((e.srcToR, e.dstToR))
.Aggregate(c => c.avg(rtt), c.max(rtt), c.min(rtt))
\end{lstlisting} \vspace{-0.1cm}

\lstset{
	basicstyle=\fontsize{8}{8}\selectfont\ttfamily,
	xleftmargin=1.0ex,
	framexleftmargin=1.0ex,
	frame=tb, 
	breaklines=true,
	captionpos=b,
	label=list:correct-spi,
	}
\begin{lstlisting}[caption={A text query for computing histogram data for per-tenant analytics job latency and resource utilization. JobStats is an object to store job-related information.}, label={lst:dnnlog-query}]
patterns={"*tenant name*", "*job running time*","*cpu util*","*memory util*"}
Stream
.Window(10_SECS)
.Map(l -> l.trim().toLowerCase())
.Filter(l -> patterns.stream().anyMatch(s->l.contains(s)))
.Map(j -> new JobStats(j.split('=')))
.Map(j -> j.stat = width_bucket(j.stat,0,100,10))
.GroupApply(j.tenant_name,j.stat_name,j.stat)
.Aggregate(c -> c.count())
\end{lstlisting}

\Paragraph{Workloads.} We use two datasets: \textbf{\pingmesh{}} dataset as
described in Section~\ref{sec:motivation} and \textbf{\dnnlog{}}, a
text-based log which includes tenant name, job running time in milliseconds
along with CPU and memory utilization for handling tenant-wise performance
issues for jobs running in an analytics cluster. We run the following
queries:

\begin{myitemize}
\item \textbf{\qone{}} (Listing~\ref{lst:motivating-query}) on
    \pingmesh{} dataset. The filter predicate delivers 14\% filter-out
    rate.

\item \textbf{\qtwo{}} (Listing~\ref{lst:tor-level-count-probe}) on
    \pingmesh{} dataset. It measures network latency
    aggregates for ToR-to-ToR pairs by joining the input stream with a table
    that maps server IP address to its ToR switch ID.

\item \textbf{\dnnlog{}} (Listing~\ref{lst:dnnlog-query}) on
    \dnnlog{} dataset. It parses unstructured logs and bucketizes per-tenant
    latency and resource utilization to create histograms.
\end{myitemize}

For \pingmesh{}, guided by~\cite{pingmesh}, we set each server to probe 20K
other servers at a time interval of 5 seconds. As a probe record is 86B, each
server generates data approximately at 2.62~Mbps. For \dnnlog{}, guided
by~\cite{chi} which reports text log data generated at 10s~PB per day across
200K \client{} nodes in a production system, we set each server to generate
0.62~MBps (or 4.96~Mbps) of log data. For experimentation purpose, we scale
up the data generation rate by 10$\times$, i.e., 26.2~Mbps for \pingmesh{} and 49.6~Mbps for \dnnlog{} per \client{} node.


\Paragraph{Network configuration.} We assume that a \server{} node would have
a network link of 10 Gbps~\cite{everflow}. For ease of experiment, we assume
this bandwidth is fairly utilized across 250 nodes (guided by conversations
with a large-scale datacenter operator) and 20 queries (guided
by~\cite{sonata}) per node, allowing 2.048~Mbps effective bandwidth per query
per \client{} node. We again scale up the obtained bandwidth by 10x to match
with data rate scaling.


\Paragraph{Baseline systems.} We compare \ours{} with the following systems:
(1) \textbf{\allspconfig{}} that runs a query entirely on \server{} (i.e.,
Gigascope~\cite{gigascope}), (2) \textbf{\allsrcconfig{}} that runs a query
entirely on \client{}, (3) \textbf{\filtersrcconfig{}} that applies static
operator-level partitioning and runs only filter operations on \client{}
(i.e., Everflow~\cite{everflow}), (4) \textbf{\bestopconfig{}} that applies a
solver to dynamically allow the best operator-level partitioning (i.e.,
Sonata~\cite{sonata}), and
(5) \textbf{\coarsedp{}} (or \textbf{LoadBalance-DP}) that applies coarse-grained
data partitioning at query level to split the input stream between \client{}
and \server{} proportional to available compute on the nodes (i.e.,
M3~\cite{m3-in-situ}).


\subsection{Query Throughput Analysis}
\label{sec:query-throughput-comparison}

\begin{figure}
    \includegraphics[height=37mm,width=\linewidth]{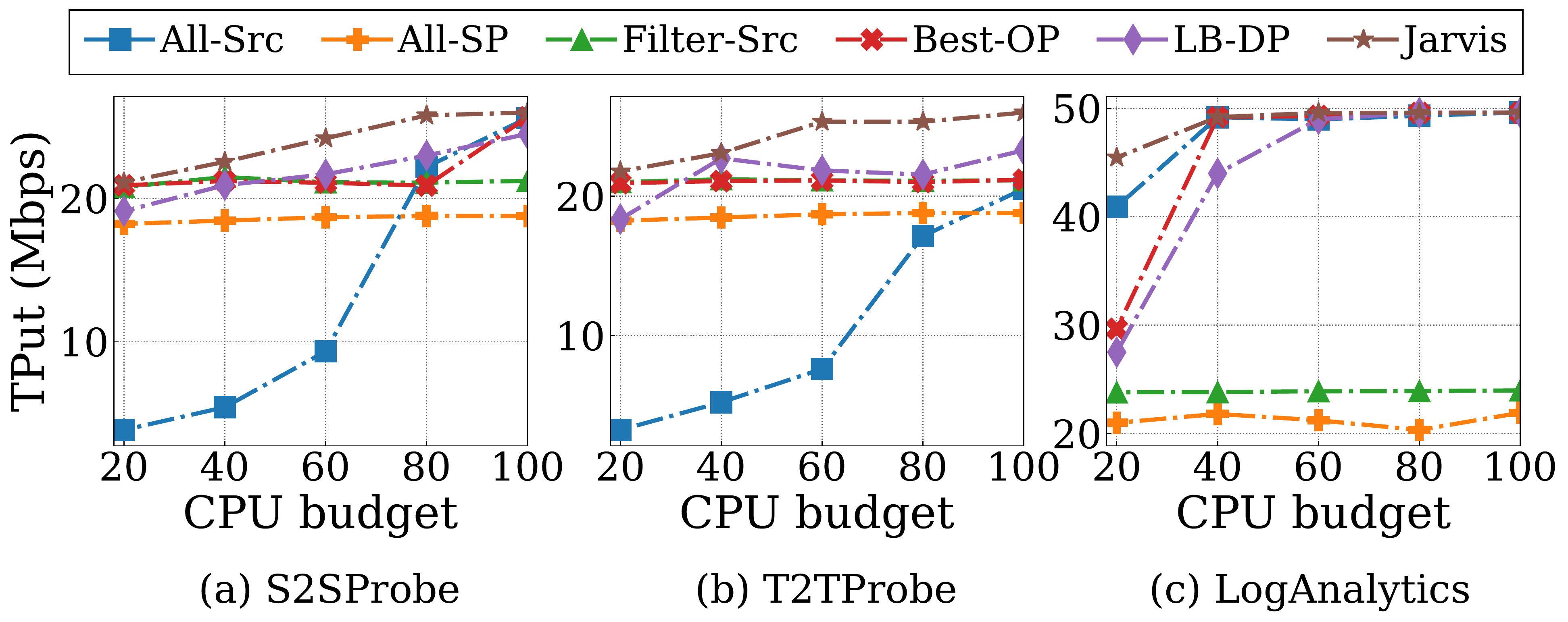}
    \caption{Query throughput over varying CPU budgets in \% of a single core. Stream joined with a table of size 500 in (b).}
    \label{fig:all-throughput}
    \vspace{-0.35cm}
\end{figure}

We use a single \client{} setup and evaluate query throughput for different
partitioning strategies while varying available compute resource on
\client{}. \ours{} incurs little overhead, consuming less than
1\% of a single core during \texttt{Profile} and \texttt{Adapt} phases.


\Paragraph{\qone{} query.} Figure~\ref{fig:all-throughput}(a) shows the 
query throughput on \qone{}.  The query requires nearly 85\% CPU
to execute entirely on \client{}. Thus, the throughput in \allsrcconfig{}
declines drastically when CPU budgets are lower than 80\%. Network bandwidth
becomes a bottleneck for \filtersrcconfig{} as F operator is not effective in
filtering out data. \bestopconfig{} executes F and G+R on \client{} only at
100\%. For lower CPU budgets, it hits compute bottleneck and runs only F at
the source as its compute cost is just 13\%. \allspconfig{} is restricted by available network bandwidth, 
and thus its throughput does not change with available CPU. \coarsedp{} generates higher 
amounts of network traffic compared to \ours{} since its goal is to balance the
compute load between \client{} and \server{} nodes. On the contrary, \ours{} partially processes the 
input of the G+R operator within available compute resource to reduce network 
traffic. \ours{} outperforms other
techniques in the 40-80\% CPU budget range, with throughput gains of 2.6$\times$ and
1.16$\times$ over \allsrcconfig{} and \coarsedp{}, respectively at 60\% CPU, and
1.25$\times$ over \bestopconfig{} at 80\% CPU.


\Paragraph{\qtwo{} query.} Figure~\ref{fig:all-throughput}(b) shows the
query throughput on \qtwo{}. This query's compute resource requirements exceed one core
due to an expensive join (J) operator. 
Thus, \allsrcconfig{} cannot handle the input rate even at 100\% CPU, resulting in significant throughput 
reduction for lower compute budgets. Both \filtersrcconfig{} and \bestopconfig{} execute only F at
the source while \bestopconfig{} cannot accommodate J operator even at 100\%
CPU. \coarsedp{} relieves compute load on \client{} but the reduction
is not sufficient for significant throughput gains. 
\ours{} performs \datapart{} to process the input partially on the J 
operator, thereby reducing network traffic, outperforming other techniques in the 40-100\% CPU range. Note that the J operator is followed by a projection on the fields, srcToR, destToR and rtt; hence, the output size of the projection is less than the input size of the J operator, leading to
data reduction.
\ours{} outperforms \allsrcconfig{} by 4.4$\times$ at 40\% CPU and \bestopconfig{} by 1.2$\times$ 
between 60-100\% CPU range. 

\Paragraph{\dnnlog{} query.} \dnnlog{} is relatively cheaper and uses 31\%
CPU to process the input at 49.6 Mbps. \allsrcconfig{} achieves lower throughput
than \ours{} at 20\% CPU as it is resource constrained. \filtersrcconfig{}
executes filtering on \client{}, but is bound by network cost due to low
filter-out rate. \bestopconfig{} can perform the filter and
map operators at the source, thus outperforming \filtersrcconfig{}.
\allspconfig{} is always bound by the network, and hence \ours{} outperforms
it by 2.3$\times$ in the 40-100\% CPU range. For 20-40\% CPU range,
\coarsedp{} transfers up to 45\% of the input over the network, resulting in
significant network bottlenecks. \ours{}' \datapart{}
reduces compute cost even at 20\% CPU budget, as shown in
Figure~\ref{fig:all-throughput}(c), outperforming \bestopconfig{}
and \coarsedp{} by 1.5$\times$.


\subsection{Convergence Analysis}
\label{sec:convergence-cost-eval}
Next, we evaluate how fast \ours{} adapts to changes in resource conditions on
\client{} by measuring the convergence time in number of epochs. We compare
\ours{} against \textbf{LP only} which runs \algo{} without fine-tuning (i.e.,
model-based approach in~\cite{heuristics-op-placement-dsps}) and \textbf{w/o
LP-init} which runs \algo{} by initializing \p{}s to zero (i.e.,
model-agnostic approach in~\cite{heuristics-op-placement-dsps}). We do not
compare against other methods that take minutes to compute a new query plan
(e.g., Sonata~\cite{sonata}).


\begin{figure}
    \includegraphics[width=\linewidth]{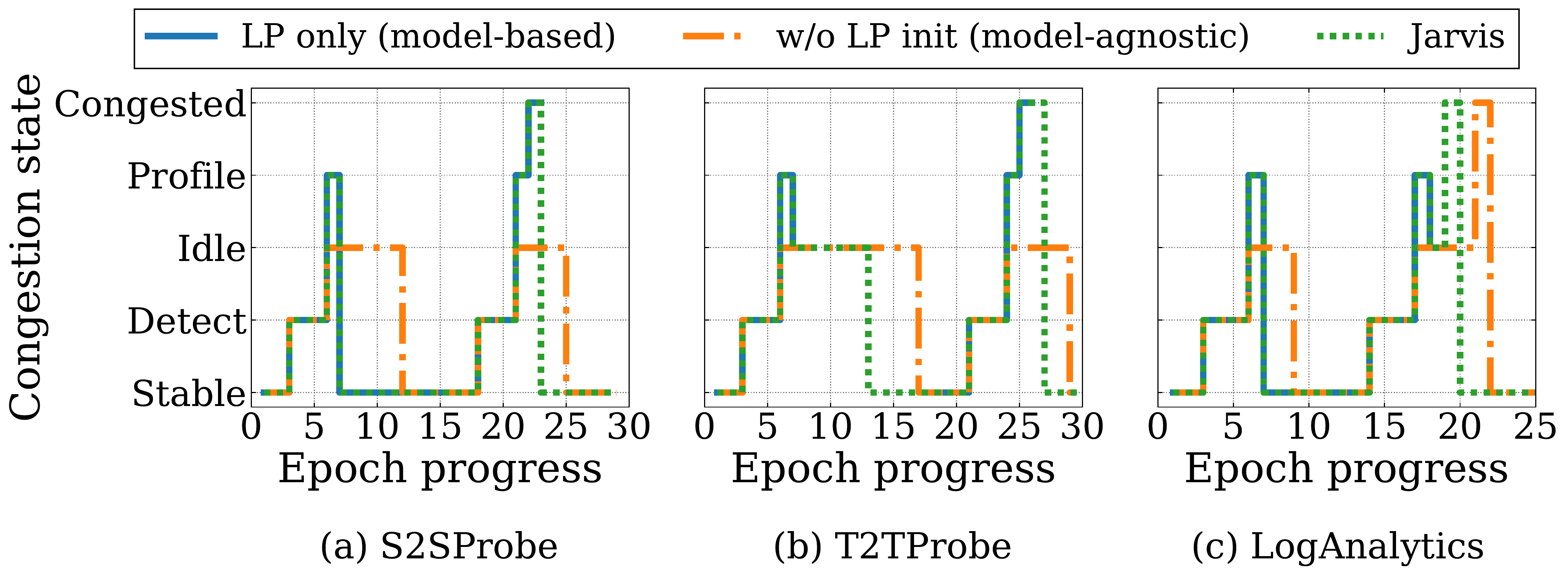}
    \caption{Convergence analysis of \ours{} compared with model-based approach and model-agnostic approach.}
    \label{fig:all-convergence}
    \vspace{-0.4cm}
\end{figure}

\Paragraph{\qone{} query.} Figure~\ref{fig:all-convergence}(a) shows the
results as we vary the available compute resource on the \client{}. Note that
three epochs are required to detect that compute budget has changed, while avoiding
triggering adaptation due to scheduling noise in the system. When the compute
budget changes at the $3^{rd}$ epoch (10\%$\rightarrow$90\% CPU), \ours{}
reduces the convergence time from six epochs in \modelagnostic{}, down to one
epoch when employing initialization using LP solver.
\modelbased{} also stabilizes the query using the output of LP
solver. When the compute budget drops at the $18^{th}$
(90\%$\rightarrow$60\% CPU), the query reaches a stable state within two and
four epochs for \ours{} and \modelagnostic{}, respectively. The additional
epoch for \ours{} is required because profiling within a one-second epoch is
not sufficient for G+R to process all records, resulting in less accurate
estimates for the cost of G+R. The inaccurate profiling also
prevents \modelbased{} from stabilizing the query.

\Paragraph{\qtwo{} query.} Performance of a join-bound query is affected by
the size of the static table. As shown in
Figure~\ref{fig:all-convergence}(b), we vary the available compute and the
size of the static table by starting with 10\% CPU and a static table of size
50, then switching to 100\% CPU in the $3^{rd}$ epoch, and finally increasing
the static table size by 10x to cause congestion.

Convergence duration is reduced from 11 epochs (in \modelagnostic{}) to seven
epochs in \ours{}, when the budget is increased to 100\%. The number of
epochs for \ours{} incurred after profiling is attributed to the fact that the
expensive J operator cannot be executed on all records to get accurate
profiling estimates. As a result, the downstream G+R operator is not profiled
accurately. Thereafter, fine-tuning plays a critical role in stabilizing the
query in \ours{}. When the table size increases, the compute cost of J
operator increases leaving no resources for G+R to execute. It takes five and three
epochs to converge in \modelagnostic{} and \ours{}, respectively.
The inaccurate profiling prevents \modelbased{} from converging for both
changes in resource conditions. Note that we manually reset load factors to stabilize 
query for the next run, at $18^{th}$ epoch.



Figure~\ref{fig:all-convergence}(c) shows the results of \dnnlog{}. We see
similar trends as observed in \qone{} and \qtwo{} queries.


\begin{figure}
\centering     
\subfigure[]{\label{fig:b}\includegraphics[height=30mm,width=36mm]{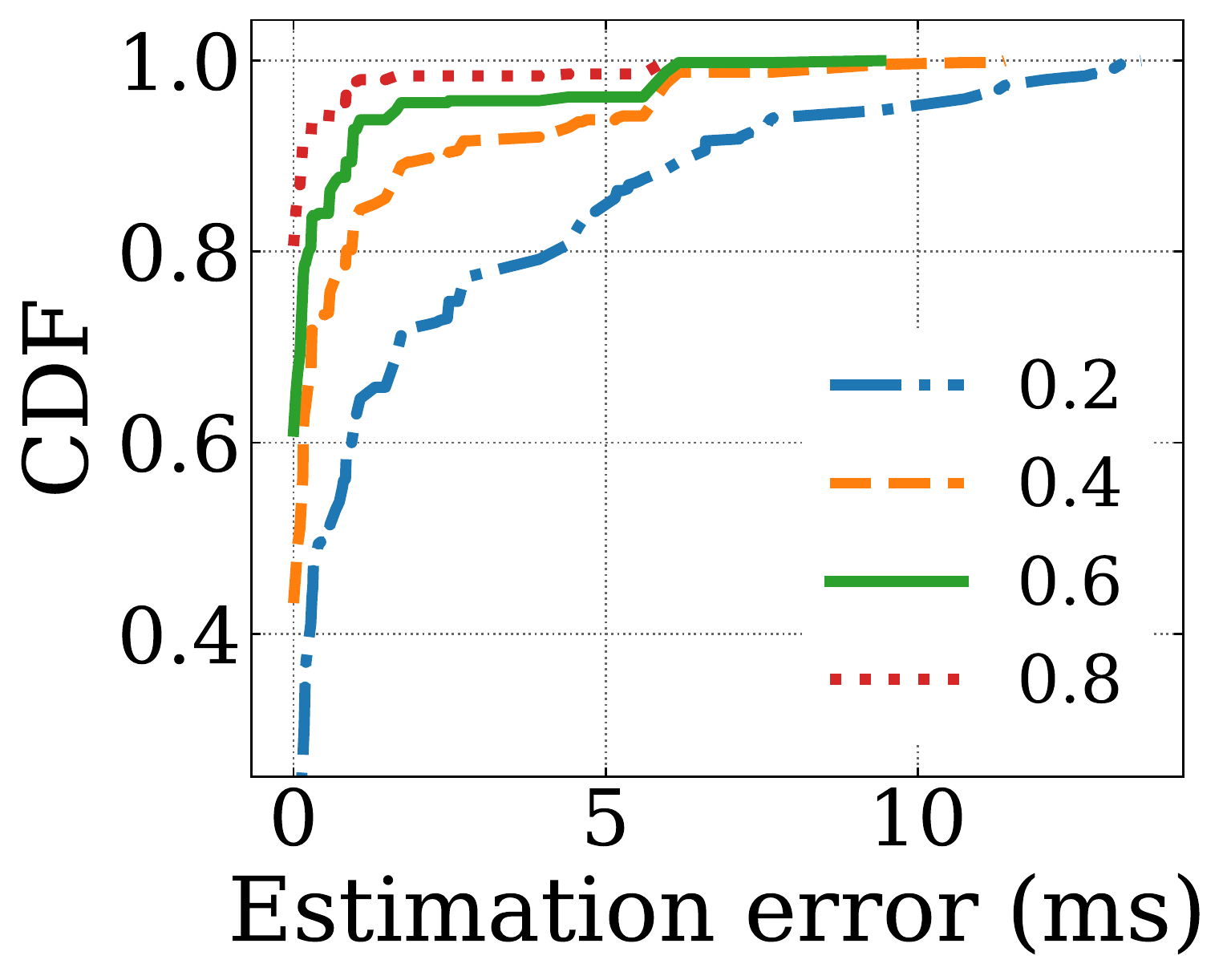}}
\subfigure[]{\label{fig:a}\includegraphics[height=30mm,width=36mm]{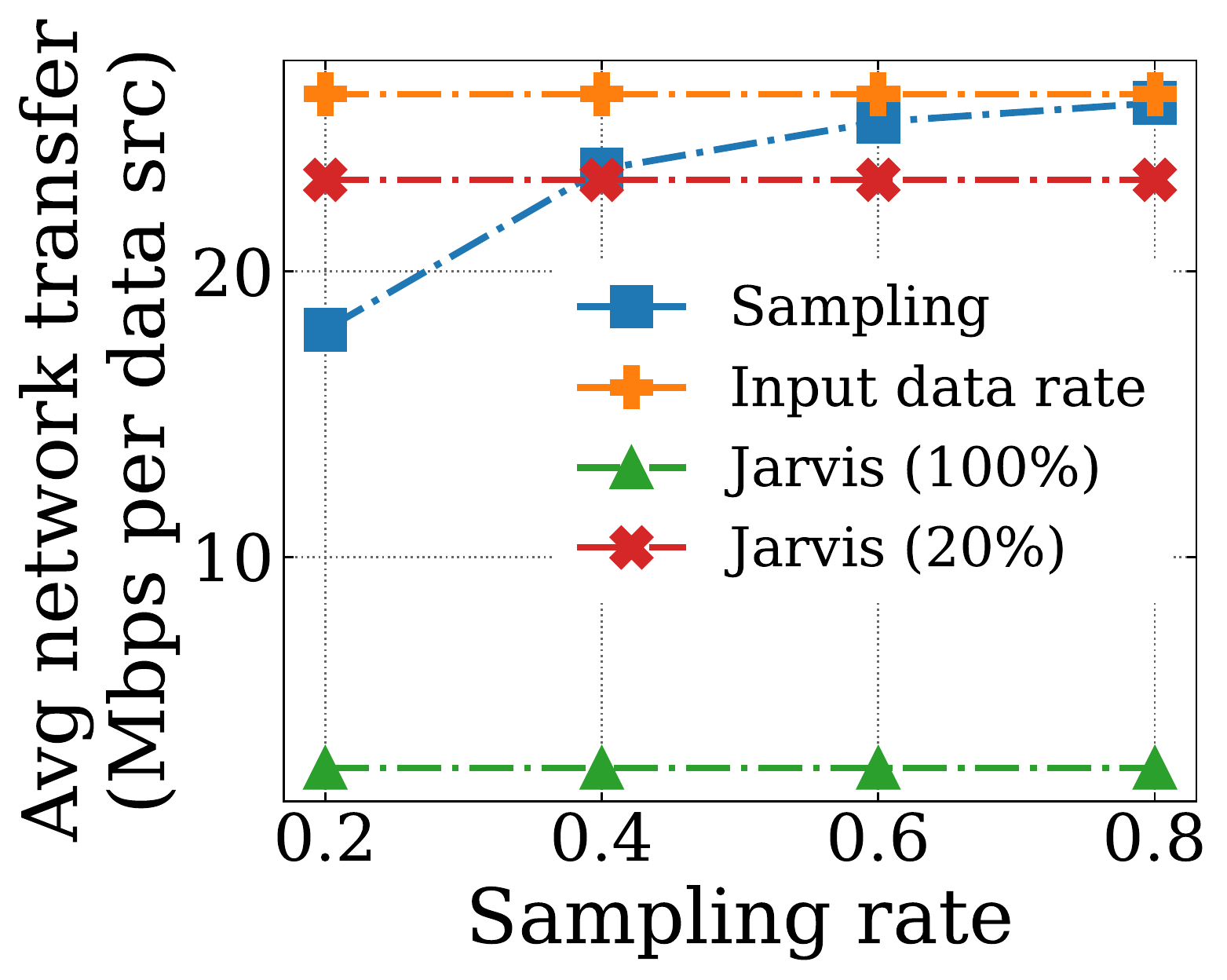}}
\caption{\label{fig:sampling_s2sprobe} (a) Cumulative distribution function (CDF) of estimation error for server probe latency using sampling. (b) Network transfer costs with 100\% and 20\% CPU budget.
}
\vspace{-0.35cm}
\end{figure}

\Paragraph{Impact of number of operators.} As we increase the number of query
operators, we expect \modelagnostic{} to
require longer time to converge. We analyze algorithm convergence via a simulator
that performs an exhaustive search of different execution configurations with
possible query resource needs and compute budgets, while measuring associated
convergence cost for each configuration. The simulator does not consider the 3 additional epochs to detect a resource change before triggering adaptation. We find that convergence time can increase to as high as 21 epochs in the worst case, with
four operators; detailed analysis can be found in~\cite{jarvis-extended}. LP solver can thus be a valuable part of design in improving convergence cost in
such cases. We do not compare against \modelbased{} and \ours{}, as the simulator does not consider the profiling estimation errors which occur in real environments, and hence the query refinement would converge within an epoch.



\subsection{Comparison against Data Synopsis}
\label{sec:sampling-vs-jarvis}
Data synopses have been proposed to reduce network transfer
costs. We quantify the potential of the window-based sampling protocol (WSP)~\cite{continuous-sampling-streams}, which constructs continuous samples from 
distributed streams in multiple \client{}s. We apply WSP to
Scenario 1 in Section~\ref{sec:background}
using the query in Listing~\ref{lst:motivating-query}. 
In this scenario, violating the search service latency SLA degrades user 
satisfaction and reduces revenues~\cite{web-search-predictive-parallel}. Developers need to quickly
correlate SLA violations with the alerts in Scenario 1 to determine if the
violation is due to a network issue~\cite{pingmesh}. This necessitates
accurate alerts to be generated. To understand the impact of data synopsis in this scenario, Figure~\ref{fig:sampling_s2sprobe} plots the error in estimating
the range of probe latencies for each server pair and the required network bandwidth to transmit the results to the \server{} while varying the WSP's sampling rate.

We observe that with 0.6 and 0.8 sampling rates, we have 85-90\% of the estimation errors within 1
ms, an acceptable error given the alert threshold of 5~ms used in Scenario 1. However, network transfer 
savings are not significant for these high sampling rates. Lower sampling rates
(i.e., 0.2 and 0.4) result in significant network bandwidth reduction (10-32\% of input rate), but 
at the cost of high estimation errors. Specifically, 20-40\% of the estimation errors exceed 1~ms for sampling rates of 0.2 and 0.4, and 20\% even exceeds 5~ms for 0.2 sampling rate. The reason for the
high errors is because high probe latencies which correspond to network issues are sparse in the dataset. When 
such probes are missed during sampling, the query can significantly underestimate the probe latencies observed 
by a node. Critically, for such low sampling rates, WSP misses 10-38\% of the alerts that should be generated given the alert threshold. Finally, we note that the network bandwidth reduction obtained by \ours{} (i.e., 11.4-90\% of input rate) is same or better, without compromising accuracy.

\subsection{Multiple Data Source Nodes}
\label{sec:scalability-analysis}
\begin{figure}
    \includegraphics[height=37mm,width=\linewidth]{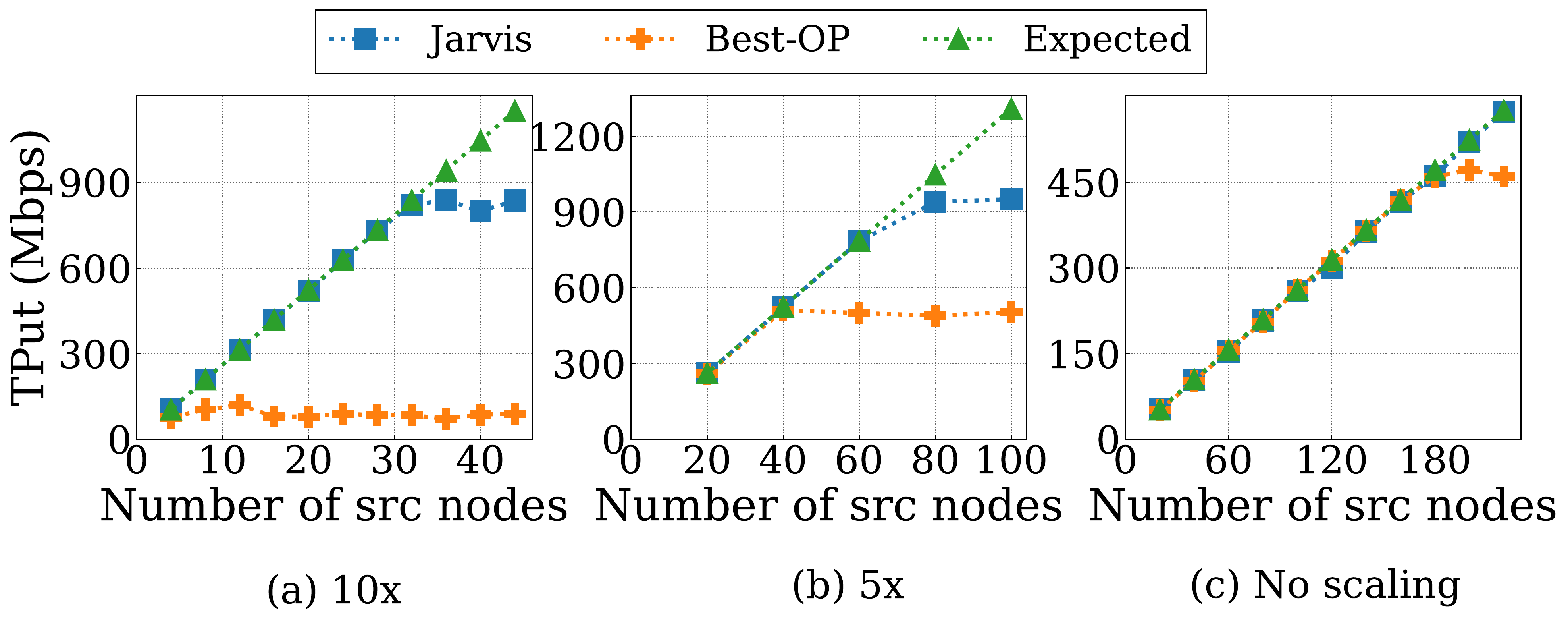}
    \caption{Query throughput over varying number of source nodes and different input rates.}
    \label{fig:scalability-throughput}
    \vspace{-0.35cm}
\end{figure}

We discuss the efficacy of \ours{} when multiple \client{}s are sending data
to a single \server{} node. We compare \ours{} against \bestopconfig{} (i.e., the
state-of-the-art in dynamic operator-level partitioning)  on \pingmesh{}'s \qone{} query
while varying the number of \client{} nodes for different input rates.

\Paragraph{Query throughput.} Figure~\ref{fig:scalability-throughput}(a) varies the number of \client{} nodes 
for an input rate of 26.2 Mbps, which is scaled by 10$\times$ over the dataset's calculated rate. On each \client{}, 
we set CPU to 55\% to ensure that \bestopconfig{} executes only the F operator while not fully utilizing 
the given CPU budget. In \bestopconfig{}, F operator does not reduce data significantly, so the policy 
suffers from network bottleneck as soon as we add more \client{}s. In contrast, \ours{} scales up to 32 
nodes without impacting throughput. Beyond 32 nodes, throughput degradation is observed due to network bottleneck.

Figure~\ref{fig:scalability-throughput}(a) varies the number of \client{} nodes for an input 
rate of 13.2 Mbps (5$\times$ scaling). On each \client{}, we set the available CPU to 30\%, to reflect the 
change in query compute demand from decreasing the input rate. \bestopconfig{} scales to 40 nodes after 
which it becomes network bottlenecked. \ours{} scales up to nearly 70 nodes, 75\% improvement in
number of \client{}s supported over \bestopconfig{}. 

Finally, when  the input rate is set to 2.62 Mbps (Figure~\ref{fig:scalability-throughput}(c)) 
and 5\% of the CPU is allocated to the query, \bestopconfig{} starts to 
degrade in throughput at 180 nodes while \ours{} is seen to scale even for 250 \client{}s.

\Paragraph{Query latency.} \ours{} improves epoch processing latency over \bestopconfig{} due to
reduced network traffic. For instance, when both policies are able to handle the input rate (e.g.,
5$\times$ scaling and 40 \client{}s), \ours{} improves median latency by 3.4$\times$ (from 1800 ms down to 500 ms) over \bestopconfig{} in the configuration. Similarly, \ours{} reduces the max latency from five seconds down to two seconds. For configurations where \bestopconfig{} cannot keep up with input rate due to network bottleneck (e.g., 5$\times$ and 60 nodes), we observe that the max latency of \bestopconfig{} grows beyond 60 seconds while \ours{} maintains it  within five seconds. 

\subsection{Multiple Queries on Data Source Node}
\label{sec:multiple-nodes-eval}
\begin{figure}
    \includegraphics[height=37mm,width=\linewidth]{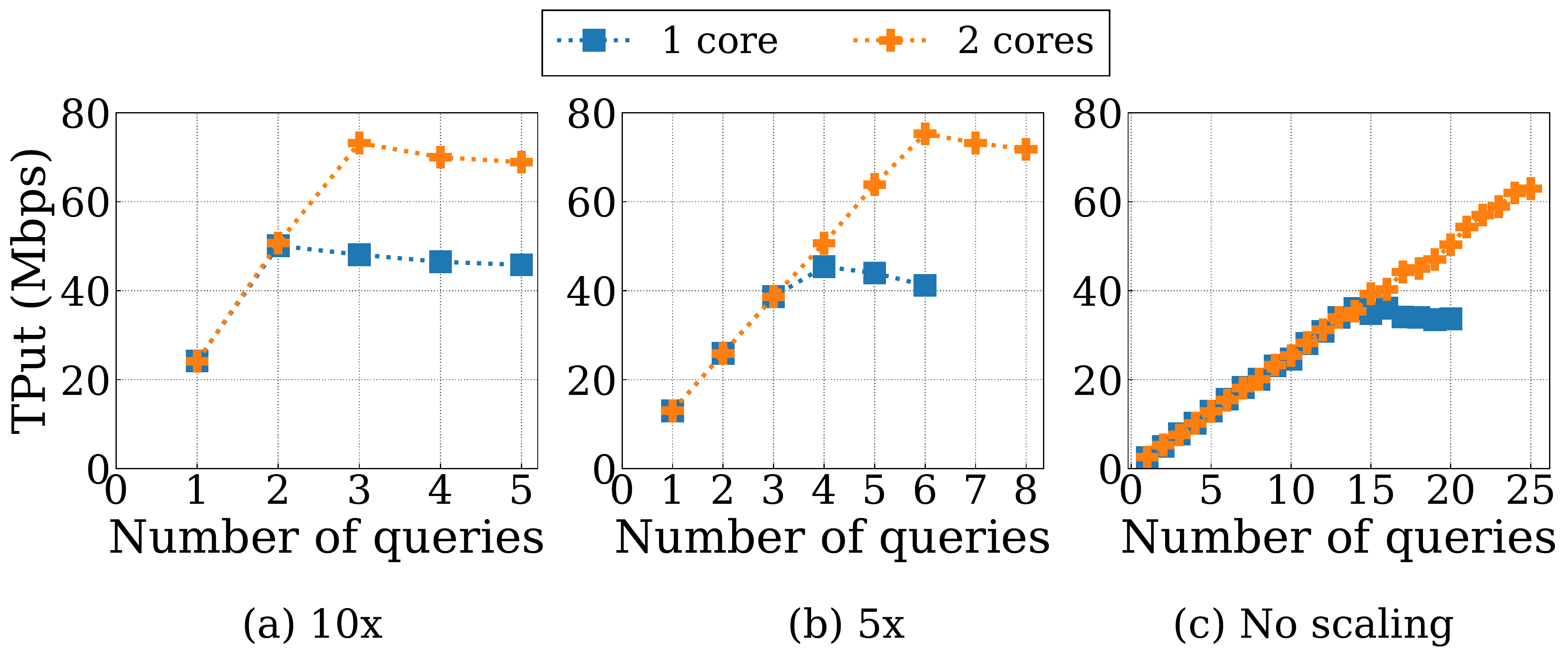}
    \caption{Query throughput for executing multiple queries on \client{} at different input rates.}
    \label{fig:multiple-queries-eval}
    \vspace{-0.35cm}
\end{figure}

Finally, we investigate the implications 
when multiple queries are executed on a \ours{}-enabled single-\client{} 
node. Our experiment runs multiple instances of \qone{} query 
while each instance is configured to use a fixed
amount of CPU resource (via fixed \p{}s).
Figure~\ref{fig:multiple-queries-eval} plots \emph{aggregate} query throughput
for various per-query input data rates and for single- and dual-core \client{}s.

We observe that there is no significant interference among query instances
until the system is bottlenecked by the compute budget. Under system stress at
10$\times$ input scaling, single-core throughput saturates at two queries given
per-query CPU demand of 55\%. Two-core throughput does not increase beyond three queries. 
At 5$\times$ scaling, per-query CPU demand drops to 30\% and \ours{} supports up to 
four and six queries on a single- and two-core setup, respectively. At no input scaling and per-query CPU demand of 5\%, \ours{} supports 15 queries and 25 queries with one and two cores, respectively.

\section{Related Work}
\label{sec:RelatedWork}
\Paragraph{In-situ analytics.} They reduce data movement by
executing query operations locally on the \client{} nodes. iMR~\cite{imr} uses
data summaries to trade off accuracy for performance.  
Rule-based heuristics~\cite{lambdaflow,lprof,scoop} statically determine which operators are pushed to \client{}; e.g.,  operators beyond the first stateful operator cannot be pushed. Such static partitioning quickly becomes sub-optimal in our system under dynamic resource conditions (Section~\ref{sec:motivation}). 

MapReduce implementations adapted 
for streaming applications~\cite{m3-in-situ, redoop} perform compute load balancing
across map operator instances by splitting input load across them (detailed comparison against \ours{} can be found in Section~\ref{sec:query-throughput-comparison}.) DIRAQ~\cite{diraq} and FlexAnalytics~\cite{flexanalytics} execute indexing and compression 
operators at the \client{}. Task-level (akin to operator-level) dynamic
resource allocation is enabled by different frameworks; for instance, Spark uses Mesos and YARN~\cite{spark-dynamic-scaling-backpressure}. Operator-level placement is evaluated extensively in Section~\ref{sec:query-throughput-comparison}.

\Paragraph{Stream operator placement.} It determines how to
place DAG operators across computing nodes for efficient query
execution. These approaches are based on operator-level partitioning~\cite{network-aware-decentralized,
decentral-tree-operator, decentral-placement-only-net, operator-tree-bokhari,
utility-op-placement-decentral, sbon}, which is less effective than \datapart{} when applied in our setup. Prior work finds an optimal placement on heterogeneous resources~\cite{opt-replicate-place-sigmetrics, opt-placement-fog} requiring solving 
an NP-hard optimization problem, which makes it impractical when fast adaptation is required across large number of \client{}s (see Section~\ref{sec:drawbacks}). Flouris et al. suggest several heuristics to
assign operators near \client{} to minimize network transmission costs under
geo-distributed sites~\cite{wan-cep-network-opt}. All the proposed algorithms consider 
coarse-grained operator-level partitioning, which is  less effective compared to 
\ours{}. Furthermore, apart from greedy heuristics, they  exhibit exponential time 
complexity in the number of \client{} nodes, 
when considering placement of each query instance's operator across 
\client{} and \server{} nodes.

Nardelli et al. study model-based and model-free heuristics
for operator placement~\cite{heuristics-op-placement-dsps}, concluding that
there is no one-size-fits-all solution. Model-based approaches have 
been implemented in centralized~\cite{network-aware-decentralized} and 
decentralized~\cite{decentral-placement-only-net} systems for operator 
placement. They rely on accurate query cost estimates, which are computationally expensive to profile on the \client{}. Cost model-agnostic approach based on rule-based heuristics has been 
studied in~\cite{decentral-tree-operator}. In Section~\ref{sec:convergence-cost-eval}, 
we show that the combination of model-based and model-agnostic approaches (as 
in \ours{}) outperforms each approach when applied individually.

\Paragraph{Parallel query processing.} Prior work improves query
performance by dynamically alleviating compute bottlenecks through operator
scaling on input data. DS2~\cite{ds2} makes a scaling decision over all
query operators at a time given the correct query performance tracing result.
Stela~\cite{per-op-stela} prioritizes scaling of operators that have the
greatest impact on overall query throughput. Similarly, other works in this
domain all make scaling decisions at operator
level~\cite{game-theory-scaling, decentral-scaling,nephele,
model-predictive-control-scaling, per-op-icdcs-autoparallel,
per-op-elastic-scale-ipdps, per-op-elastic-scale-tpds,
per-op-elastic-iot-cloud, per-op-predictable-low-lat-iot, dhalion}. Applying
these approaches to our setup requires that the query operators be
replicated between each \client{} and its parent \server{} node. However,
compute resource allocation on the \client{} in our system happens at the
query level; and a query consists of multiple operators with different data
reduction capabilities. Our optimization goal is thus different: incoming
records need to be carefully apportioned across multiple co-located query
operators, in order to minimize data transfer costs within a compute budget.

Conventional key-based load splitting
strategies~\cite{key-splitting-pipeflow, key-splitting-graphcep, key-split-debs-near-optimal,
key-split-massive-scaleout, key-split-millwheel, key-split-s4,
key-split-vldb-holistic-view} are
complementary to our work. These strategies assign input key ranges to
different stateful operator instances such that each instance keeps the state
of a distinct, non-overlapping key range. Integrating key-based load
splitting into \ours{} may further reduce network transfer costs by minimizing the number of output keys sent from the \client{}. Recent studies investigate assigning subsequences of records or windows to operator
instances based on the currently assigned compute load in the host
node~\cite{local-load-sliding-window-harness,
local-load-online-shuffle-grouping, local-load-elastic-ppq}. These approaches
do not assume that compute resources in each \server{} node are shared by
multiple \client{}s (see Figure~\ref{fig:arch-monitoring-scale}). Therefore,
compute load on the \server{} node depends on jointly considering splitting decisions over \client{} nodes. This is exponential in the number
of \client{}s.

SkewTune~\cite{straggler-skewtune} and Google
Dataflow~\cite{straggler-google-dataflow} employ work-stealing techniques for
straggler mitigation while others~\cite{per-op-debs-window-cep,
local-load-power-of-both-choices} use local estimates of compute load to
detect a congested host node and re-route input records to a different node
in a greedy manner. In our setup, each \client{} has a fixed compute budget
allocated for a query consisting of multiple operators with different data
reduction capabilities. We could consider re-routing input records to a \server{} node, when an operator does not result in a significant data reduction. This boils down to ``\modelagnostic{}'' which is evaluated for convergence speed in Section~\ref{sec:convergence-cost-eval}.
To facilitate throughput comparison of \ours{} with compute load balancing techniques at the query level, we evaluate ``\coarsedp{}'' in Section~\ref{sec:query-throughput-comparison}.

\Paragraph{Stream computations on the edge.} EdgeWise~\cite{edgewise} is a
streaming engine run on the edge, improving throughput and latency by
prioritizing query operators experiencing backpressure.
Droplet~\cite{droplet} and Costless~\cite{costless} address the operator
placement problem between edge and cloud resources for video analytics
applications. They neither leverage the benefits of the data-level
partitioning nor respond to dynamic resource conditions on the edge. 
\ours{} does both.



\section{Conclusion}
We presented \ours{}, a fully decentralized data-level query partitioning engine for server monitoring systems. Our analysis using real-world monitoring query workloads suggests that \ours{} substantially improves system throughput while quickly adapting to changes in resource conditions. 
 
\section{Acknowledgements}
We thank the anonymous reviewers, Ruoyu Sun and Niao He from UIUC for their feedback and comments. This work was supported in part by the National Science Foundation under Grant No. SHF 1617401, and in part by the Laboratory Directed Research and Development program at Sandia National Laboratories, a multi-mission laboratory managed and operated by National Technology and Engineering Solutions of Sandia, LLC, a wholly owned subsidiary of Honeywell International, Inc., for the U.S. Department of Energy’s National Nuclear Security Administration under contract DE-NA0003525, the 2018 Research Fund (1.180079.01) of UNIST(Ulsan National Institute of Science \& Technology), and Samsung Data \& Information Technology Center.

\bibliographystyle{IEEEtran}
\bibliography{jarvis_dist}
\end{document}